\documentclass[12pt,sort&compress]{elsarticle}
\usepackage[utf8]{inputenc}

\usepackage{bm}

\usepackage{mathrsfs}
\usepackage{amsmath}
\usepackage{amsfonts}
\usepackage{amsthm}
\usepackage{color}
\usepackage{sansmath}

\usepackage{array}

\usepackage{caption}
\usepackage{subcaption}

\usepackage{float}
\usepackage{url}
\usepackage{multicol}
\usepackage[top=1in,bottom=1in,left=1in,right=1in]{geometry}
\usepackage[pdfborder={0 0 0},colorlinks,allcolors=blue]{hyperref}
\usepackage{txfonts}
\usepackage{setspace}
\usepackage{arydshln}
\usepackage{enumitem}
\usepackage{lipsum}
\usepackage{siunitx,booktabs}
\usepackage{nth}
\usepackage{accents} 

\theoremstyle{definition}
\newtheorem{remark}{Remark}

\allowdisplaybreaks

\begin{document}
\begin{frontmatter}

\title{Treatment of near-incompressibility and volumetric locking in higher order material point methods}

\author[sbu]{Ram Mohan~Telikicherla}

\author[sbu,iacs]{Georgios~Moutsanidis\corref{cor1}}
\ead{georgios.moutsanidis@stonybrook.edu}
\cortext[cor1]{Corresponding author}

\address[sbu]{Department of Civil Engineering, Stony Brook University, Stony Brook, NY 11794, USA}

\address[iacs]{Institute for Advanced Computational Science, Stony Brook, NY 11794, USA}

\journal{Computer Methods in Applied Mechanics and Engineering}

\begin{abstract}
We propose a novel projection method to treat near-incompressibility and volumetric locking in small- and large-deformation elasticity and plasticity within the context of higher order material point methods. The material point method is well known to exhibit volumetric locking due to the presence of large numbers of material points per element that are used to decrease the quadrature error. Although there has been considerable research on the treatment of near-incompressibility in the traditional material point method, the issue has not been studied in depth for higher order material point methods. Using the $\overline{\mathbf{B}}$ and $\overline{\mathbf{F}}$ methods as our point of departure we develop an appropriate projection technique for material point methods that use higher order shape functions for the background discretization. The approach is based on the projection of the dilatational part of the appropriate strain rate measure onto a lower dimensional approximation space, according to the traditional $\overline{\mathbf{B}}$ and $\overline{\mathbf{F}}$ techniques, but tailored to the material point method.  The presented numerical examples exhibit reduced stress oscillations and are free of volumetric locking and hourglassing phenomena.
\end{abstract}


\end{frontmatter}

\section{Introduction}
\label{sec:intro}

The material point method (MPM) \cite{sulsky1994particle} is a computational technique that emerged from the particle-in-cell (PIC) method \cite{harlow1962particle,evans1957particle}. It combines Lagrangian particles with an Eulerian grid. A continuum body is discretized by a set of Lagrangian particles, also referred to as material points, that are free to move on top of a background Eulerian mesh in a non-conforming manner. The governing equations are posed in a variational (weak) form, while the material points carry history dependent quantities and serve as quadrature points for the weak form of the governing equations. The weak form is solved on the background Eulerian mesh and the solution is then communicated back to the particles and the material quantities are updated. MPM is ideal for problems involving large deformations and extreme material distortion, where traditional mesh based computational techniques, such as the finite element method (FEM) \cite{hughes2012finite}, fail due to mesh entanglement and the need for frequent mesh update or remeshing \cite{sulsky1993particle,sulsky1993particle1,york2000fluid,stomakhin2013material,andersen2010modelling}. Since the initial development of the method there has been considerable research in the field and the interested reader should consult \cite{bardenhagen2004generalized,chen2002evaluation,zhang2008material,zhang2011material,sadeghirad2011convected,yerro2015material,soga2016trends,homel2017field,kumar2017modelling,tielen2017high,gan2018enhancement,moutsanidis2019modeling,kumar2019scalable,moutsanidis2020iga,de2020material} and the references therein.

However, MPM, like the majority of traditional variational-based numerical methods, is known to exhibit volumetric locking when it comes to small and large deformation nearly incompressible elasticity and plasticity. This is a result of the presence of large numbers of material points per background cell that leads to overconstrained systems \cite{chen2018vp}. A significant number of material points per cell is required to reduce the integration error, therefore lowering the number of material points is not a viable solution to locking. Although treatment of incompressibility and volumetric locking have been extensive areas of research in conventional mesh-based computational methods \cite{hughes1977equivalence,malkus1978mixed,hughes1980generalization,belytschko1984hourglass,simo1990class,neto996design,klaas1999stabilized,kasper2000mixed,neto2005f,elguedj2008b,brezzi2012mixed}, they are still open research subjects in MPM. MPM researchers have mostly relied on operator splitting schemes where a divergence-free velocity field is solved for \cite{zhang2017incompressible,kularathna2017implicit,zhang2018augmented}, and on mixed formulations \cite{love2006energy,mast2012mitigating,chen2018vp,iaconeta2019stabilized}. Both approaches reduce pressure oscillations and alleviate locking. However, they are fairly complex to implement, introduce additional variables, and lead to increased computational times. The authors in \cite{coombs2018overcoming} overcame these difficulties by developing an equivalent $\overline{\textbf{F}}$ method \cite{neto996design,neto2005f,elguedj2008b} for MPM without the complexities associated with operator splitting or multiple field variational formulations. The method was based on the projection of the dilatational (volumetric) part of the deformation gradient tensor onto a space of constant functions. 

All the above-mentioned works primarily focused on the conventional MPM where $C^0$-continuous linear shape functions are used to discretize the background domain. Recently, with the advent of Isogeometric Analysis (IGA) \cite{hughes2005isogeometric}, the MPM community has started employing higher order spline basis functions for the background discretization \cite{steffen2008analysis,tielen2017high,gan2018enhancement,de2021extension,bazilevs2017new,moutsanidis2018hyperbolic,moutsanidis2020iga}. The use of higher order and smooth functions results in a continuous representation of the strain rate, and therefore prevents jumps in the stress and other history variables as the material points cross the element boundaries. Thus, the so-called cell-crossing instability \cite{bardenhagen2004generalized} is eliminated. Past experience has also shown that the use of higher order basis functions reduces the effects of near-incompressibility to a certain extent. Nevertheless, the use of higher order basis functions alone is not enough to entirely eliminate volumetric locking. 

Thus, in this work, we develop a novel $\overline{\mathbf{F}}$ approach tailored to higher order material point methods such as the B-spline MPM (BSMPM) \cite{gan2018enhancement} and the isogeometric material point method (IGA-MPM) \cite{moutsanidis2020iga}. The proposed methodology is based on the projection of the dilatational part of the velocity gradient onto a lower dimensional space. The numerical examples show that the framework results in reduced stress oscillations, and is free of volumetric locking or low energy modes.

The present paper is organized as follows. In Sections \ref{sec:IGA} and \ref{sec:MPM} we briefly present the basics of IGA and MPM, respectively. In Section \ref{sec:bbar} we present the basics of $\overline{\mathbf{B}}$ and $\overline{\mathbf{F}}$ methods. In Section \ref{sec:framework} we outline the proposed framework. In Section \ref{sec:examples} we present the supporting numerical examples, and in Section \ref{sec:conclusions} we provide concluding remarks.

\section{Isogeometric analysis}
\label{sec:IGA}

Higher order MPM was inspired from IGA based on B-splines and Non-Uniform Rational B-splines (NURBS). In this section, to be self-contained, we briefly recall the basics of IGA \cite{hughes2005isogeometric} based on B-splines. B-splines exhibit excellent mathematical properties, such as derivative-continuity across element boundaries, optimal approximation \cite{bazilevs2006isogeometric1}, and the ability to be refined through knot insertion and degree elevation. 

\subsection{B-splines}

A necessary component for the construction of B-splines is the knot vector. A knot vector in 1D is a non-decreasing set of coordinates in the parametric domain written as $ \Xi= $ $\{\xi_{1}, \xi_{2} , \dots, \xi_{n+p+1} \}$, where $ \xi_{i} \in \mathbb{R} $ is the $i$\textsuperscript{th} knot, $i$ is the knot index, $i = 1,2,\dots,n+p+1$, $p$ is the polynomial order, and $n$ is the number of B-spline basis functions. Knots divide the parametric domain into elements. 

For a given knot vector, the B-spline basis functions are defined recursively starting with piecewise constants ($p=0$):
\begin{equation}
N_{i,0} (\xi) = \begin{cases} 1 \; \text{if} \; \xi_{i} \leqslant \xi < \xi_{i+1}, \\ 0 \; \text{otherwise.} \end{cases} \label{eq:1}
\end{equation}
For $p = 1,2,3,\dots,$ they are defined by
\begin{equation}
\label{eq:2}
N_{i,p}(\xi) = \frac{\xi - \xi_{i}}{\xi_{i+p}-\xi_{i}} N_{i,p-1}(\xi) + \frac{\xi_{i+p+1}-\xi}{\xi_{i+p+1}-\xi_{i+1}} N_{i+1,p-1}(\xi),
\end{equation}
which is the Cox-de Boor recursion formula \cite{cox1972numerical}.

Knot vectors may be open or closed. In an open knot vector the first and last knot values appear $p+1$ times. B-spline basis functions constructed using an open knot vector are interpolatory at the endpoints of the parametric interval, which facilitates imposition of boundary conditions. In general, B-splines are not interpolatory at interior knots. Only open knot vectors are employed in the present work.

Basis functions of order $p$ have $p-m_{i}$ continuous derivatives at knot $\xi_{i}$, where $m_{i}$ is the multiplicity of the knot $\xi_{i}$ in the knot vector. The B-spline basis functions are pointwise non-negative, satisfy the partition of unity, that is,
\begin{equation}
\label{eq:3}
\sum_{i=1}^{n} N_{i,p}(\xi)=1~~\forall \xi \in \Xi, 
\end{equation}
and the support of each basis function $N_{i,p}$ is compact and contained in the interval $[\xi_{i},\xi_{i+p+1}]$.

\subsection{Analysis framework}

We write $\hat{N}(\pmb{\xi})$ and $N(\mathbf{x})$ to refer to a generic B-spline basis function defined on the parametric and physical domains, respectively. We also make use of a single-index notation, and let indices $A,B,C,\dots$ label the basis functions. In this setting, the geometry mapping may be expressed as
\begin{equation}
\label{eq:4}
\mathbf{x}(\pmb{\xi}) = \sum_{A=1}^{\rm{n_{np}}} \mathbf{x}_{A} \hat{N}_{A}(\pmb{\xi}),
\end{equation}
where $\rm{n_{np}}$ denotes the number of control points in the mesh with coordinates given by $\mathbf{x}_{A}$'s. This mapping may be restricted to a patch or element.

The IGA solution in the parametric domain, taken to be scalar-valued for the purposes of illustration, is assumed to be governed by the same B-spline basis functions, and may be expressed as
\begin{equation}
\label{eq:5}
\hat{u}^h(\pmb{\xi}) = \sum_{A=1}^{\rm{n_{np}}} u_{A} \hat{N}_{A}(\pmb{\xi}),
\end{equation}
where $u_{A}$'s are the control variables or degrees of freedom (DOF). The IGA solution in the physical domain is defined as a push-forward of its parametric counterpart given in (\ref{eq:5}) by the geometrical mapping given by (\ref{eq:4}), and may be expressed as
\begin{equation}
\label{eq:6}
u^h(\mathbf{x}) = \sum_{A=1}^{\rm{n_{np}}} u_{A} N_{A}(\mathbf{x}), 
\end{equation}
where
\begin{equation}
\label{eq:7}
N_{A}(\mathbf{x}) = \hat{N}_A(\pmb{\xi}^{-1} (\mathbf{x})).
\end{equation}
Equations (\ref{eq:4})-(\ref{eq:7}) constitute the well-known isoparametric construction widely used in FEM and IGA. The above construction guarantees optimal approximation properties of B-splines spaces as shown in \cite{bazilevs2006isogeometric1}. The first and second partial derivatives of the basis functions in (\ref{eq:7}) with respect to physical coordinates, are computed using the chain rule in a manner similar to FEM.

\section{The material point method}
\label{sec:MPM}

In this section we first recall the basics of the conventional explicit material point method, and we then describe the implementation of higher order MPM.

\subsection{Governing equations}

We consider the strong-form of the linear momentum balance for a continuum object expressed in an updated Lagrangian frame as: Given $\mathbf{b} : \Omega_t \rightarrow \mathbb{R}^{3}, \mathbf{g}_{V} : \Gamma_{g} \rightarrow \mathbb{R}^{3}, \mathbf{h} : \Gamma_{h} \rightarrow \mathbb{R}^{3}, \text{find} \; \mathbf{v}: \Omega_t \rightarrow \mathbb{R}^{3} $ such that:

\begin{equation}
\label{eq:momentum}
\rho \, \dot{\mathbf{v}} = \nabla\cdot\bm{\sigma} + \rho \, \mathbf{b}~~~\text{at every }\mathbf{x}\in\Omega_t\text{ ,}
\end{equation}
\begin{equation}
\label{eq:bc1}
(\bm{\sigma} \cdot \mathbf{n}) \, |_{\Gamma_h} = \mathbf{h},
\end{equation}
\begin{equation}
\label{eq:bc2}
\mathbf{v} \, |_{\Gamma_g} = \mathbf{g}_{V},
\end{equation}
where $\nabla$ is the vector differential operator, $\rho$ is the mass density, $\mathbf{v}$ is the velocity of the material, $\bm{\sigma}$ is the Cauchy stress, $\mathbf{b}$ is the body force per unit mass, $\Omega_t$ is a time-dependent region deforming with velocity $\mathbf{v}$, 
$\mathbf{h}$ is the prescribed traction on boundary $\Gamma_h$, $\mathbf{g}_{V}$ is the prescribed velocity on boundary $\Gamma_g$, $\mathbf{n}$ is the unit outward normal to the boundary, and the superimposed dot denotes time differentiation. The weak form corresponding to the problem defined in \eqref{eq:momentum} - \eqref{eq:bc2} is: Find $\mathbf{v} \in \mathcal{L}$, such that for all $\mathbf{w} \in \mathcal{V}$,

\begin{equation}
\label{eq:weak}
\int_{\Omega_t} \mathbf{w} \rho \dot{\mathbf{v}} \, d \Omega \; + \int_{\Omega_t} \nabla \mathbf{w} : \bm{\sigma} \, d \Omega \; - \int_{\Omega_t} \mathbf{w} \rho \mathbf{b} \, d \Omega - \int_{\Gamma_h} \mathbf{w} \mathbf{h} \, d \Gamma = \mathbf{0} \, ,
\end{equation}
where $\mathcal{L} = \big\{ \mathbf{v} \; | \; \mathbf{v} \in H^1 (\Omega), \mathbf{v}|_{\Gamma_{g}} = \mathbf{g}_{V} \big\}$, and
$\mathcal{V} = \big\{ \mathbf{w} \; | \; \mathbf{w} \in H^1 (\Omega), \mathbf{w}|_{\Gamma_{g}} = 0 \big\}$, are the corresponding trial and test function spaces, respectively. As mentioned in Section \ref{sec:intro}, the weak form \eqref{eq:weak} is solved on a background Eulerian grid, whereas the Lagrangian material points are used to discretize ${\Omega_t}$ and serve as integration points for \eqref{eq:weak}.

\subsection{MPM solution procedure}

Within the context of MPM various formulations have been adopted to solve \eqref{eq:weak}. In the numerical examples presented in this paper we use the so-called ``modified update stress last"(MUSL) \cite{de2020material}, which is presented hereafter. The following notation is used: subscripts $i$ and $mp$ indicate a background grid node and a material point, respectively. $m_i$ and $\mathbf{v}_i$ are the mass and velocity vector at grid node $i$, respectively, $m_{mp}$ and $\mathbf{v}_{mp}$ are the mass and velocity vector at material point $mp$, respectively, $N_i (\mathbf{x}_{mp}$) is the shape function of grid node $i$ evaluated at material point $mp$ with position vector $\mathbf{x}_{mp}$, $V_{mp}$ is the volume of material point $mp$ in the current configuration, $\Delta t$ is the time step, and $n$ refers to the $n$-th time step. The MPM computation at $(n+1)$-th time step proceeds as follows:
\begin{enumerate}
\item Map the mass from the material points to the corresponding grid nodes
\begin{equation}
m_{i}^{n} = \sum\limits_{mp} N_{i} (\mathbf{x}_{mp}^{n}) \, m_{mp},
\end{equation}
where $\sum \limits_{mp}$ denotes summation over all particles that are supported in $N_{i}$.
\item Map the momentum from the material points to the corresponding grid nodes
\begin{equation}
(m \mathbf{v})_{i}^{n} = \sum\limits_{mp} N_{i} (\mathbf{x}_{mp}^{n}) \, m_{mp} \, \mathbf{v}_{mp}^n.
\end{equation}
\item Obtain the corresponding grid velocity $\mathbf{v}_{i}$
\begin{equation}
\mathbf{v}_{i}^{n} = \frac{(m \mathbf{v})_{i}^n}{m_{i}^n}.
\end{equation}
\item Compute the internal force vector, $\mathbf{f}_{i}^{int}$, at grid nodes
\begin{equation}
(\mathbf{f}_{i}^{int})^{n} = - \sum \limits_{mp} \nabla N_{i} (\mathbf{x}_{mp}^{n}) \, \bm{\sigma}_{mp}^{n} V_{mp}.
\end{equation}
\item Compute the external force vector, $\mathbf{f}_{i}^{ext}$, at grid nodes
\begin{equation}
(\mathbf{f}_{i}^{ext})^{n} = \sum \limits_{mp} N_{i} (\mathbf{x}_{mp}^{n}) \, m_{mp} \, \mathbf{b} \, (\mathbf{x}_{mp}^{n}).
\end{equation}
\item Apply essential boundary conditions and solve for the grid acceleration $\mathbf{a}_{i}$
\begin{equation}
\mathbf{a}_{i}^{n} = \frac{(\mathbf{f}_{i}^{int})^{n} + (\mathbf{f}_{i}^{ext})^{n}}{m_{i}^{n}}.
\end{equation}
\item Update the grid velocity
\begin{equation}
\mathbf{v}_{i}^{n+1} = \mathbf{v}_{i}^{n} + \mathbf{a}_{i}^{n} \, \Delta t.
\end{equation}
\item Update the particle velocity and position
\begin{equation}
\mathbf{v}_{mp}^{n+1} = \mathbf{v}_{mp}^{n} + \Delta t \sum\limits_{i} N_{i} (\mathbf{x}_{mp}^{n}) \, \mathbf{a}_{i}^{n},
\end{equation}
\begin{equation}
\mathbf{x}_{mp}^{n+1} = \mathbf{x}_{mp}^{n} + \Delta t \sum\limits_{i} N_{i} (\mathbf{x}_{mp}^{n}) \, \mathbf{v}_{i}^{n+1},
\end{equation}
where $\sum \limits_{i}$ denotes summation over all grid nodes whose basis function supports the particle with coordinates $\mathbf{x}_{mp}$.
\item Map the updated particle momentum to the grid nodes to get the updated grid momentum $(m \overline{\mathbf{v}})_{i}$
\begin{equation}
(m \overline{\mathbf{v}})_{i}^{n+1} = \sum\limits_{mp} N_{i} (\mathbf{x}_{mp}^{n}) \, m_{mp} \, \mathbf{v}_{mp}^{n+1}.
\end{equation}
\item Apply essential boundary conditions and obtain the corresponding updated grid velocity $\overline{\mathbf{v}}_{i}$
\begin{equation}
\overline{\mathbf{v}}_{i}^{n+1} = \frac{(m \overline{\mathbf{v}})_{i}^{n+1}}{m_{i}^n}.
\end{equation}
\item Compute the particle velocity gradient
\begin{equation}
\nabla \mathbf{v}_{mp}^{\,{n+1}} = \sum\limits_{i} \nabla N_{i} (\mathbf{x}_{mp}^{n}) \overline{\mathbf{v}}_{i}^{n+1}.
\end{equation}
\item Update the particle deformation gradient $\mathbf{F}_{mp}$
\begin{equation}
\mathbf{F}_{mp}^{n+1} = (\mathbf{I} + \nabla \mathbf{v}_{mp}^{\,{n+1}} \Delta t ) \, \mathbf{F}_{mp}^{n}.
\end{equation}
\item Update the particle volume
\begin{equation}
V_{mp}^{n+1} = | \mathbf{F}_{mp}^{n+1} | V_{mp}^{0}.
\end{equation}
\item Compute the particle rate of deformation tensor $\mathbf{D}_{mp}$ and spin tensor $\pmb{\omega}_{mp}$
\begin{equation}
\label{eq:Drate}
\mathbf{D}_{mp}^{n+1} = \frac{1}{2} ( \nabla \mathbf{v}_{mp}^{\,{n+1}} + \nabla \mathbf{v}_{mp}^{\,T,{n+1}} ),
\end{equation}
\begin{equation}
\pmb{\omega}_{mp}^{n+1} = \frac{1}{2} ( \nabla \mathbf{v}_{mp}^{\,{n+1}} - \nabla \mathbf{v}_{mp}^{\,T,{n+1}} ).
\end{equation}
\item Update the particle Cauchy stress $\bm{\sigma}_{mp}$ through the corresponding constitutive model. For elasto-plastic materials, the constitutive response is typically defined in rate form, relating an objective Cauchy stress rate (e.g. Jaumann rate $\bm{\sigma}^{\nabla J} = \dot{\bm{\sigma}} - \bm{\sigma} \pmb{\omega}^{T} - \pmb{\omega} \bm{\sigma}$) to its power conjugate strain rate measure. In this work, the corresponding stain rate measure is the rate of deformation tensor defined in terms of the velocity gradient as seen in in \eqref{eq:Drate}.
\end{enumerate}

\subsection{Higher order material point methods}
The implementation of higher order MPM is straightforward and has minor differences with the conventional MPM framework. The main difference is the use of control points and control variables instead of nodes and nodal variables. The control mesh is a result of a tensor product grid and the control points are in 1-to-1 correspondence to the B-spline basis functions as defined in \eqref{eq:2}. The rest of the framework follows easily from the conventional MPM implementation where the $C^0$-continuous linear shape functions are now replaced with higher order B-spline functions. For example, in the previously presented MPM solution procedure, $N_i \, (\mathbf{x}_{mp})$ corresponds to the B-spline basis function as defined in \eqref{eq:2}, where $i$ refers to a control point on the background mesh.

\begin{remark}
For the problems considered in this work, essential boundary conditions are only applied at the boundaries of the background domain. As mentioned in Section \ref{sec:IGA}, open knot vectors are used. Therefore, the basis functions are interpolatory at the background domain boundaries, and the imposition of essential boundary conditions is practically the same as in the traditional MPM.
\end{remark}

\section{$\overline{\mathbf{B}}$ and $\overline{\mathbf{F}}$ methods}
\label{sec:bbar}

\subsection{The $\overline{\mathbf{B}}$ method}

In this section we summarize the basics of the $\overline{\mathbf{B}}$ and $\overline{\mathbf{F}}$ methods following \cite{elguedj2008b}. We start by defining the boundary value problem of linear elastostatics:
Given $\mathbf{f} : \Omega \rightarrow \mathbb{R}^{3}, \mathbf{g}_{D} : \Gamma_{g} \rightarrow \mathbb{R}^{3}, \mathbf{h} : \Gamma_{h} \rightarrow \mathbb{R}^{3}, \text{find} \; \mathbf{u}: \Omega \rightarrow \mathbb{R}^{3} $ such that:

\begin{equation}
\nabla \cdot \bm{\sigma} + \mathbf{f} = \bm{0} ~~~\text{at every }\mathbf{x}\in\Omega\text{ ,}
\end{equation}

\begin{equation}
(\bm{\sigma} \cdot \mathbf{n}) \, |_{\Gamma_h} = \mathbf{h},
\end{equation}
\begin{equation}
\mathbf{u} \, |_{\Gamma_g} = \mathbf{g}_{D},
\end{equation}
where $\mathbf{u}$ is the displacement, and $\mathbf{g}_{D}$ is the prescribed displacement on $\Gamma_{g}$.

Next we define the strain tensor as:
\begin{equation}
\pmb{\varepsilon} = \nabla^{s} \mathbf{u},
\end{equation}
where $\nabla^{s}$ is the symmetric gradient operator, and the stress tensor through the generalized Hooke's law as:
\begin{equation}
\bm{\sigma} = \pmb{\mathbb{C}} : \pmb{\varepsilon} 
\end{equation}
where $\pmb{\mathbb{C}}$ is the tensor that contains the elastic coefficients. In index notation, these coefficients are defined in terms of the so-called Lame parameters $\lambda$ and $\mu$.
\begin{equation}
\mathbb{C}_{ijkl} = \lambda \, \delta_{ij} \, \delta_{kl} + \mu \, ( \delta_{ik} \, \delta_{jl} + \delta_{il} \, \delta_{jk})
\end{equation}
where
\begin{equation}
\lambda = \frac{2 \mu \nu}{(1-2\nu)}
\end{equation}
\begin{equation}
\mu = \frac{E}{2(1+\nu)}
\end{equation}
and $\nu$ is the Poisson's ratio and $E$ is the Young's modulus. Thus, the stress is
\begin{equation}
    \pmb{\sigma} = \pmb{\mathbb{C}} : \pmb{\varepsilon} = \lambda \, \text{tr} (\pmb{\varepsilon}) \mathbf{I} \, + 2\mu \, \pmb{\varepsilon},
\end{equation}
where $\mathbf{I}$ is the identity tensor.

In the near incompressible case, $\nu \rightarrow 0.5$ and as a result $\lambda \rightarrow \infty$. For the pure incompressible case, the constitutive equation is expressed as:
\begin{equation}
\bm{\sigma} = - p \mathbf{I} + 2\mu \pmb{\varepsilon},
\end{equation}
where $p$ is the hydrostatic pressure, and an additional equation that describes the kinematic condition of incompressibility must be added to the original boundary value problem,
\begin{equation}
\nabla \cdot \mathbf{u} = \bm{0}.
\end{equation}
In order to achieve a purely displacement based formulation though, we consider the \textbf{nearly} incompressible regime, meaning that the ratio $\lambda$ / $\mu$ is large but not infinite. 

The basic idea of the $\overline{\mathbf{B}}$ is to split the strain tensor into its deviatoric and volumetric (dilatational) parts as:
\begin{equation}
\pmb{\varepsilon} (\mathbf{u}) = \pmb{\varepsilon}^{dev} (\mathbf{u}) + \pmb{\varepsilon}^{dil} (\mathbf{u})
\end{equation}
where
\begin{equation}
\pmb{\varepsilon}^{dil} (\mathbf{u}) = \frac{1}{3} (\nabla \cdot \mathbf{u}) \, \mathbf{I}.
\end{equation}
To achieve a formulation that is suitable for the nearly incompressible case, the dilatational part of the strain tensor is replaced by a projected one.
\begin{equation}
\overline{\pmb{\varepsilon}}^{dil} (\mathbf{u}) = \pi (\pmb{\varepsilon}^{dil} (\mathbf{u}))
\end{equation}
where $\pi$ is a linear projection operator.
Extending this operation to the strain-displacement matrix $\mathbf{B}$ we have
\begin{equation}
\overline{\mathbf{B}} = \mathbf{B}^{dev} + \overline{\mathbf{B}} \, ^{dil} 
\end{equation}
and
\begin{equation}
\mathbf{B} = \mathbf{B}^{dev} + \mathbf{B}^{dil} 
\end{equation}
Starting from a minimum potential energy principle and following the approach presented in \cite{elguedj2008b} we arrive at a $\overline{\mathbf{B}}$ variational formulation stated as:

Find $\mathbf{u} \in L$, such that for all $\mathbf{w} \in V$
\begin{equation}
\overline{a} (\mathbf{w},\mathbf{u}) = (\mathbf{w},\mathbf{f}) + (\mathbf{w},\mathbf{h})_{\Gamma_{h}}
\end{equation}
where
\begin{equation}
\overline{a} (\mathbf{w},\mathbf{u}) = \int_{\Omega} \overline{\pmb{\varepsilon}} (\mathbf{w}) \, \mathbb{C} \, \overline{\pmb{\varepsilon}} (\mathbf{u}) \, d \Omega
\end{equation}
\begin{equation}
(\mathbf{w},\mathbf{f}) = \int_{\Omega} \mathbf{w} \cdot \mathbf{f} \, d\Omega
\end{equation}
\begin{equation}
(\mathbf{w},\mathbf{h})_{\Gamma_{h}} = \int_{\Gamma_{h}} \mathbf{w} \cdot \mathbf{h}\,  d\Gamma
\end{equation}
and $L = \big\{ \mathbf{u} \; | \; \mathbf{u} \in H^1 (\Omega), \mathbf{u}|_{\Gamma_{g}} = \mathbf{g}_{D} \big\}$,
$V = \big\{ \mathbf{w} \; | \; \mathbf{w} \in H^1 (\Omega), \mathbf{w}|_{\Gamma_{g}} = 0 \big\}$.

\subsection{The $\overline{\mathbf{F}}$ method}\label{sec:fbar_background}

In the finite strain regime, the boundary value problem for a body with reference configuration $\mathcal{B}$ is defined as: Given body force $\bm{\mathcal{F}} : \mathcal{B} \rightarrow \mathbb{R}^{3}$, prescribed displacement $\mathbf{g}_{D}: \partial_{u}
\mathcal{B} \rightarrow \mathbb{R}^{3}$, and prescribed Piola traction $\bm{\mathcal{H}} : \partial_{\mathcal{H}} \mathcal{B} \rightarrow \mathbb{R}^{3}$, find $\mathbf{u} \in \mathcal{L}$ such that

\begin{equation}
\nabla^{\bm{X}} \cdot \mathbf{P} + \bm{\mathcal{F}} = \bm{0} ~~~\text{at every }\mathbf{x}\in \mathcal{B},
\end{equation}
\begin{equation}
\mathbf{u} \, |_{\partial_{u} \mathcal{B}} = \mathbf{g}_{D},
\end{equation}
\begin{equation}
\mathbf{P} \cdot \mathbf{N} \, |_{\partial_{\mathcal{H}} \mathcal{B}} = \bm{\mathcal{H}},
\end{equation}
where $\mathbf{N}$ is the unit exterior normal on $\partial \mathcal{B}$, and $\mathcal{L}$ is the space of admissible deformations defined by 
\begin{equation}
\mathcal{L} = \big\{ \mathbf{u} : \Omega \rightarrow \mathbb{R}^{3} \, | \, \mathbf{u} \in H^1 (\mathcal{B}) \, | \, \text{det} (\mathbf{I} + \nabla^{\mathbf{X}} \mathbf{u}) > 0, \mathbf{u} |_{\partial_{u} \mathcal{B}} = \mathbf{g}_{D} \big \}.
\end{equation}
Introducing a mapping $\bm{\phi} : \mathcal{B} \rightarrow \mathcal{B}'$ that takes a point in the reference configuration $\mathbf{X} \in \mathcal{B}$ to a point in the current configuration $\mathbf{x} = \bm{\phi} (\mathbf{X}) \in \mathcal{B}'$, the deformation gradient is defined as
\begin{equation}
\mathbf{F} (\mathbf{X}) = \nabla^{\bm{X}} \bm{\phi} (\mathbf{X}) = \frac{\partial \bm{\phi} (\mathbf{X})}{\partial \mathbf{X}} = \mathbf{I} + \frac{\partial \mathbf{u} (\mathbf{X})}{\partial \mathbf{X}}.
\end{equation}
$\mathbf{P}$ is the first Piola-Kirchhoff stress defined as
\begin{equation}
\mathbf{P} = \mathbf{F} \mathbf{S},
\end{equation}
where $\mathbf{S}$ is the second Piola-Kirchhoff stress
\begin{equation}
\mathbf{S} = 2\,\frac{\partial \Psi (\mathbf{C})}{\partial \mathbf{C}},
\end{equation}
$\mathbf{C} = \mathbf{F}^{T} \mathbf{F}$ is the Cauchy-Green tensor, and $\Psi$ is the free energy function.  The Cauchy stress tensor can be determined by the relation
\begin{equation}
\bm{\sigma} = \frac{1}{J} \mathbf{F} \mathbf{S} \mathbf{F}^{T}.
\end{equation}
The $\overline{\mathbf{F}}$ method is the extension of the $\overline{\mathbf{B}}$ method to the finite deformation regime. In the finite deformation case, the deformations are measured through the deformation gradient $\mathbf{F}$. In contrast to the small deformation case, the split here is multiplicative.
\begin{equation}
\label{eq:f_multiplicative}
\mathbf{F} = \mathbf{F}^{dil} \, \mathbf{F}^{dev},
\end{equation}
where $\mathbf{F}^{dil}$ and $\mathbf{F}^{dev}$ correspond to the dilatational (volumetric) and deviatoric (volume preserving) components of the deformation gradient respectively, and 

\begin{equation}
\text{det} \, \mathbf{F} = J = \text{det} \, \mathbf{F}^{dil},
\end{equation}

\begin{equation}
\text{det} \, \mathbf{F}^{dev} = 1,
\end{equation}

\begin{equation}
\mathbf{F}^{dev} = J^{-1/3} \mathbf{F},
\end{equation}

\begin{equation}
\mathbf{F}^{dil} = J^{1/3} \mathbf{I}.
\end{equation}
As was the case previously, we replace the dilatational component of the deformation gradient by an improved (projected) contribution and the new deformation gradient is defined as:

\begin{equation}
\label{eq:Fbar}
\overline{\mathbf{F}} = \overline{\mathbf{F}} \, ^{dil} \, \mathbf{F}^{dev},
\end{equation}
where
\begin{equation}
\label{eq:f_projection}
\overline{\mathbf{F}} \, ^{dil} = \pi ( \mathbf{F}^{dil} ) = \overline{J^{1/3}} \, \mathbf{I},
\end{equation}
with $\pi$ being the same projection defined in the linear case. Following the steps in \cite{elguedj2008b} we arrive at the variational problem: Find $\mathbf{u} \in \mathcal{L}$, such that $\forall \mathbf{w} \in \mathcal{V}_{u}$
\begin{equation}
\overline{a} (\mathbf{w},\mathbf{u}) = (\mathbf{w},\mathbf{f}) + (\mathbf{w},\mathbf{h})_{\partial_{h} \mathcal{B}'},
\end{equation}
where 
\begin{equation}
\overline{a} (\mathbf{w},\mathbf{u}) = \int_{\mathcal{B}'} \overline{\bm{\sigma}} (\mathbf{u}) \overline{\pmb{\varepsilon}} (\mathbf{w}) \, d v,
\end{equation}
\begin{equation}
(\mathbf{w},\mathbf{f}) = \int_{\mathcal{B}'} \mathbf{w} \cdot \mathbf{f} \, d v,
\end{equation}
\begin{equation}
(\mathbf{w},\mathbf{h})_{\partial_{h} \mathcal{B}'} = \int_{\partial_{h} \mathcal{B}'} \mathbf{w} \cdot \, \mathbf{h} \, d \gamma,
\end{equation}
\begin{equation}
\overline{\mathbf{S}} = \frac{\partial \Psi (\mathbf{E})}{\partial \mathbf{E}} (\overline{\mathbf{E}}(\mathbf{u})),
\end{equation}
\begin{equation}
\overline{\mathbf{E}} = \frac{1}{2} (\overline{\mathbf{F}}^{T} \overline{\mathbf{F}} - \mathbf{I}),
\end{equation}
\begin{equation}
\overline{\bm{\sigma}} = \frac{1}{J} \overline{\mathbf{F}} \overline{\mathbf{S}} \overline{\mathbf{F}}^{T},
\end{equation}
where $\mathbf{f}$ and $\mathbf{h}$ are the spatial counterparts of $\bm{\mathcal{F}}$ and $\bm{\mathcal{H}}$ respectively, and
$\mathcal{V}_{u} = \big\{ \mathbf{w} \; | \; \mathbf{w} \in H^1 (B), \mathbf{w} |_{\partial_{u} \mathcal{B}} = 0 \big\}$.  For elasto-plastic materials, $\overline{\mathbf{F}}$ is used in determining the modified objective Cauchy stress rate and its corresponding power conjugate strain rate measure.

In the following section we present the proposed $\overline{\mathbf{F}}$ approach for higher order material point methods, along with the specifics of the projection operator and the approximation spaces.

\section{Proposed framework}
\label{sec:framework}

As can become evident through the solution procedure outlined in Section \ref{sec:MPM}, the velocity gradient of the material point, $\nabla \mathbf{v}$, is the corresponding strain rate measure that drives the stress update. Therefore, the main idea in our approach is to derive an appropriate velocity gradient compatible with the $\overline{\mathbf{F}}$ approach. We use \cite{elguedj2008b} as our point of departure and follow the steps outlined in \eqref{eq:f_multiplicative} - \eqref{eq:Fbar}.
We then rewrite \eqref{eq:Fbar} as:
\begin{equation}
\label{eq:Fbar1}
\overline{\mathbf{F}} = \alpha \mathbf{F},
\end{equation}
where
\begin{equation}
\label{eq:Fbar2}
\alpha = \overline{J^{1/3}}{J^{-1/3}},
\end{equation}
and
\begin{equation}
\overline{J} = \pi (J).
\end{equation}
By differentiating (\ref{eq:Fbar1}) we compute
\begin{equation}
\dot{\overline{\mathbf{F}}} = \dot{\alpha} \mathbf{F} + \alpha \dot{\mathbf{F}}.
\end{equation}
Using the generating equation for the deformation gradient
\begin{equation}
\dot{\mathbf{F}} = \nabla \mathbf{v} \, \mathbf{F},
\end{equation}
we have
\begin{equation}
\label{eq:Fbar3}
\dot{\overline{\mathbf{F}}} = \dot{\alpha} \frac{\overline{\mathbf{F}}}{\alpha} + \alpha \nabla \mathbf{v} \, \mathbf{F} = \frac{\dot{\alpha}}{\alpha} \overline{\mathbf{F}} + \nabla \mathbf{v} \, \overline{\mathbf{F}}.
\end{equation}
By differentiating (\ref{eq:Fbar2}) we compute
\begin{equation}
\dot{\alpha} = \dot{\overline{J^{1/3}}}{J^{-1/3}} - \overline{J^{1/3}} (\dot{J^{1/3}}){J^{-2/3}},
\end{equation}
\begin{equation}
\label{eq:Jdot1}
(\dot{J^{1/3}}) = \frac{1}{3} J^{-2/3} \dot{J},
\end{equation}
\begin{equation}
\label{eq:Jdot2}
\dot{J} = J \, \nabla \cdot \mathbf{v}.
\end{equation}
From (\ref{eq:Jdot1}) and (\ref{eq:Jdot2}) it follows that
\begin{equation}
(\dot{J^{1/3}}) = \frac{1}{3} J^{1/3} \nabla \cdot \mathbf{v}.
\end{equation}
Then,
\begin{equation}
\label{eq:Fbar4}
\frac{\dot{\alpha}}{\alpha} = \frac{\dot{\overline{J^{1/3}}} J^{-1/3} - \frac{1}{3} \overline{J^{1/3}} J^{-1/3} \nabla \cdot \mathbf{v}}{\overline{J^{1/3}} J^{-1/3}} = \frac{\dot{\overline{J^{1/3}}}}{\overline{J^{1/3}}} - \frac{1}{3} \nabla \cdot \mathbf{v},
\end{equation}
and
\begin{equation}
\dot{\overline{J^{1/3}}} = (\dot{\pi (J^{1/3})}) = \pi (\dot{J^{1/3}}) = \frac{1}{3} \pi (J^{1/3} \nabla \cdot \mathbf{v}),
\end{equation}
where, in the latter expression, the projection and time differentiation commute. As a result, (\ref{eq:Fbar4}) becomes
\begin{equation}
\frac{\dot{\alpha}}{\alpha} = \frac{1}{3} \Bigg ( \frac{\pi (J^{1/3} \nabla \cdot \mathbf{v})}{\pi (J^{1/3})} - \nabla \cdot \mathbf{v} \Bigg ).
\end{equation}
Putting together, we can rewrite (\ref{eq:Fbar3}) as
\begin{equation}
\label{eq:Fbar5}
\dot{\overline{\mathbf{F}}} = \Bigg ( \nabla \mathbf{v} + \frac{1}{3} \Bigg ( \frac{\pi (J^{1/3} \nabla \cdot \mathbf{v})}{\pi (J^{1/3})} - \nabla \cdot \mathbf{v} \Bigg ) \Bigg ) \, \overline{\mathbf{F}}.
\end{equation}
(\ref{eq:Fbar5}) is in the form
\begin{equation}
\dot{\overline{\mathbf{F}}} = \overline{\nabla \mathbf{v}} \, \overline{\mathbf{F}},
\end{equation}
and therefore we have
\begin{equation}
\label{eq:nablau}
\overline{\nabla \mathbf{v}} = \nabla \mathbf{v} + \frac{1}{3} \Bigg( \frac{\pi(J^{1/3} \nabla \cdot \mathbf{v})}{\pi(J^{1/3})} - \nabla \cdot \mathbf{v} \Bigg) \mathbf{I}.
\end{equation}
As will become apparent through the definition of the projection operation, $\pi$ is local. Therefore, in the vicinity of the projection $\pi$ we assume very small spatial variations of the determinant $J$, and we thus eliminate it from the formulation to obtain a simpler and easier to work with formula for the velocity gradient
\begin{equation}
\label{eq:nablau_original}
\overline{\nabla \mathbf{v}} = \nabla \mathbf{v} + \frac{1}{3} \Big( \pi (\nabla \cdot \mathbf{v}) - \nabla \cdot \mathbf{v} \Big) \mathbf{I}.
\end{equation}
Using $\overline{\nabla \mathbf{v}}$ we compute $\overline{\bm{\sigma}}$ through an appropriate constitutive law (Steps 14 and 15 in the presented MPM algorithm). Following the rationale of \cite{simo2006computational}[Section 4.4] we construct 
\begin{equation}
\overline{\overline{\bm{\sigma}}} = \overline{\bm{\sigma}} + \Big( \pi (\overline{\sigma}^{\,dil}) - \overline{\sigma}^{\, dil} \Big) \mathbf{I},
\end{equation}
where $\overline{\sigma}^{\,dil}$ denotes the dilatational part of the stress (hydrostatic). The variational form previously defined in \eqref{eq:weak} now becomes: Find $\mathbf{v} \in \mathcal{L}$, such that for all $\mathbf{w} \in \mathcal{V}$, 
\begin{equation}
\label{eq:weak2}
\int_{\Omega_t} \mathbf{w} \rho \dot{\mathbf{v}} \, d \Omega \; + \int_{\Omega_t} \nabla \mathbf{w} : \overline{\overline{\bm{\sigma}}} \, d \Omega \; - \int_{\Omega_t} \mathbf{w} \rho \mathbf{b} \, d \Omega - \int_{\Gamma_h} \mathbf{w} \mathbf{h} \, d \Gamma = \mathbf{0}.
\end{equation}
The projection $\pi$ is defined in the following way
\begin{equation}
\label{eq:projection}
\pi (\nabla \cdot \mathbf{v})_{j} = \frac{\int_{\Omega_{e}} \widetilde{N}_{j} \, \nabla \cdot \mathbf{v} \, d \Omega_{e}}{\int_{\Omega_{e}} \widetilde{N}_{j} \, d \Omega_{e}} = \frac{ \sum_{mp} \widetilde{N}_{j} (\mathbf{x}_{mp}) \, \nabla \cdot \mathbf{v} (\mathbf{x}_{mp}) V_{mp}}{ \sum_{mp} \widetilde{N}_{j} (\mathbf{x}_{mp}) \, V_{mp}},
\end{equation}
\begin{equation}
\label{eq:stress_projection}
\pi (\overline{\sigma}^{\,dil})_{j} = \frac{\int_{\Omega_{e}} \widetilde{N}_{j} \, \overline{\sigma}^{\, dil} \, d \Omega_{e}}{\int_{\Omega_{e}} \widetilde{N}_{j} \, d \Omega_{e}} = \frac{ \sum_{mp} \widetilde{N}_{j} (\mathbf{x}_{mp}) \, \overline{\sigma}^{\, dil} (\mathbf{x}_{mp}) V_{mp}}{ \sum_{mp} \widetilde{N}_{j} (\mathbf{x}_{mp}) \, V_{mp}},
\end{equation}
where $\widetilde{N}$ refers to a lower order shape function, and $j$ refers to a node or control point of a corresponding lower dimensional background grid. This projection corresponds to a lumped $L^2$ projection. The reconstruction of the projected quantities on the material points is easily done as
\begin{equation}
\pi (\nabla \cdot \mathbf{v})_{mp} = \sum \limits_{j} \widetilde{N}_{j} (\mathbf{x}_{mp}) \, \pi (\nabla \cdot \mathbf{v})_{j},
\end{equation}
\begin{equation}
\pi (\overline{\sigma}^{\,dil})_{mp} = \sum \limits_{j} \widetilde{N}_{j} (\mathbf{x}_{mp}) \, \pi (\overline{\sigma}^{\, dil})_{j}.
\end{equation}

\begin{remark}
As can be seen from \eqref{eq:projection} and \eqref{eq:stress_projection} , the projection corresponding to a background control point $j$ is defined through a summation over all the material points that are supported within the basis function of control point $j$. The size of this support is $p+1$, where $p$ is the polynomial order of the basis functions. In that sense, we can say that the projection is ``local" and is defined over a small region. Thus, we assume very small variations of the determinant $J$ within the vicinity of a background control point ($J \approx \text{const}$), and we eliminate it from the projection operation in \eqref{eq:nablau}.
\end{remark}

\begin{remark}
We would like to emphasize on the difference between $N$ and $\widetilde{N}$. $N$ refers to background shape functions employed to solve the weak form of the governing equations, whereas $\widetilde{N}$ refers to functions that define the space onto which the proposed $\overline{\mathbf{F}}$ projection takes place. Therefore, it follows that two background grids are required; the first one is the grid spanned by $N_i$ and is used to solve the weak form, and the second one is the grid spanned by $\widetilde{N}_j$ and defines the space onto which the projection is performed. We take $\widetilde{N}$ to be one order lower than $N$ to reduce the number of incompressibility constraints.
\end{remark}

\begin{remark}
The efficiency of the proposed framework in treating near-incompressibility and volumetric locking can be assessed through the method of constraint counting \cite{hughes2012finite}, which is very similar to the corresponding method in mesh-based numerical methods. According to \cite{hughes2012finite}[Chapter 4], the constraint ratio $r$ is defined as
\begin{equation}
r = \frac{n_{eq}}{n_{c}},
\end{equation}
where $n_{eq}$ is the number of element equations, $n_{c}$ is the number of constraints, and the optimal $r$ value is equal to the number of spatial dimensions $n_{SD}$. A value of $r$ that is less than $n_{SD}$ leads to locking. For reduced and selective integration, and $\overline{\mathbf{B}}$ / $\overline{\mathbf{F}}$ approaches, $n_{c} = \text{min} (\text{number of independent monomials in} \; \nabla \cdot \mathbf{v}, \text{number of quadrature points})$. In MPM, where a sufficiently large number of material points is used in order to improve the quadrature quality, the number of independent monomials present in $\nabla \cdot \mathbf{v}$ always governs. For example, if quadratic B-spline shape functions are used in 2D, then $n_{eq} = 8$, $n_{c} = 8$, and $r = \frac{8}{8} = 1$, which leads to locking. In contrast, if our proposed approach is employed, $\nabla \cdot \mathbf{v}$ is projected onto a space of constants, and therefore $n_{c} = 3$, and $r = \frac{8}{3}$, which leads to a system that overcomes locking.
\end{remark}

\begin{remark}
In our approach, as presented in Section \ref{sec:MPM}, we are employing a hypoelastic material model whose evolution is driven by the velocity gradient. Therefore, the corresponding strain rate measure is the velocity gradient and the proposed projection technique takes place for the velocity gradient (i.e. equations \eqref{eq:nablau_original} and \eqref{eq:projection}). Clearly, when using different types of material models driven by quantities other than the velocity gradient, the projection needs to be applied to the appropriate strain rate or deformation measure.
\end{remark}

\begin{remark}
In this work, when it comes to higher order background discretizations, only B-splines have been considered. However, curved background domains can be constructed and discretized through the use of NURBS for the background basis functions (for details see \cite{moutsanidis2020iga}). The proposed projection framework can be easily applied to that case.
\end{remark}

\section{Numerical examples}
\label{sec:examples}

In this section we present a collection of challenging problems that demonstrate our framework's capability to treat near-incompressibility and overcome volumetric locking. Due to the explicit nature of the MPM solution procedure outlined in Section \ref{sec:MPM}, for all the presented examples, dynamic analysis is performed.

\subsection{Vibrating bar}

We would first like to assess how the proposed framework affects the convergence rate of higher order MPM. Thus, we test our formulation on a vibrating bar problem that has been studied by various researchers in the context of spline-based MPM \cite{tielen2017high,gan2018enhancement,moutsanidis2020iga,wobbes2021comparison}. A bar of length L is fixed at both ends. An initial longitudinal velocity of $v(x) = v_{o} sin \Big(\frac{\pi x}{L} \Big)$ is applied to the entire bar to introduce vibration. The following parameters are used: $L = 25$~m, Young’s modulus $E = 100$~Pa, density $\rho = 1$~kg/m$^3$, and $v_{o} = 0.1$~m/s. In the transverse direction the bar has a width of $5$~m. The dimensions of the background grid are $25$~m $\times$ $5$~m $\times$ $1$~m. Three discretizations, labeled M$N$, at refinement levels $N=1,2,3$, are considered. M1 has $12 \times 2 \times 1$ background elements and $768$ material points. M2 has $25 \times 5 \times 1$ background elements and $4,000$ material points. M3 has $50 \times 10 \times 1$ background elements and $16,000$ material points. These discretizations correspond to 4 material points per background cell per $xy$-plane direction, whereas 2 material points are employed in the out of plane $z$-direction. The problem is solved with a three dimensional MPM code and plane strain conditions are enforced by restraining any out of plane action. For the background discretization we employ $C^1$-continuous quadratic B-spline shape functions with and without or proposed projection technique. The time step used with discretization M$N$ is $\Delta t = 2 \times 10^{-4}\times 2^{1-N}$~s and the final time is $0.5$~s. The analytical solution for the displacement at time t is given as
\begin{equation}
u(x,t) = \frac{v_{o} L}{\pi c} sin \bigg(\frac{\pi c}{L} t \bigg) sin \bigg(\frac{\pi x}{L} \bigg),
\end{equation}
where $c = \sqrt{\frac{E}{\rho}}$ is the elastic wave speed.

\begin{figure}[!htbp]\centering
\includegraphics[width=0.9\textwidth]{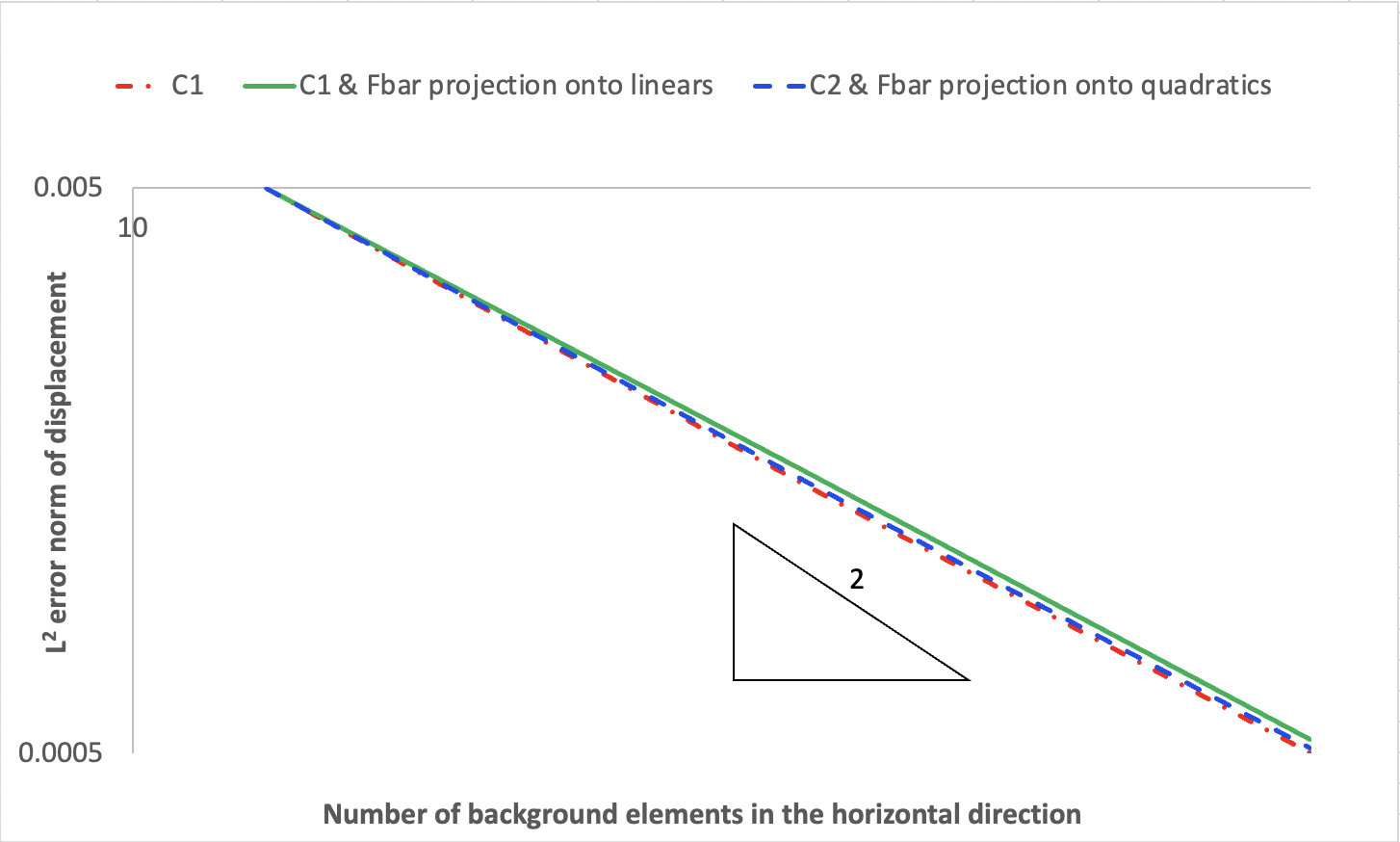}
\caption{Vibrating bar. Convergence of the L$^2$ norm of the displacement error at 0.5 s}
\label{fig:convergence}
\end{figure}

Figure \ref{fig:convergence} shows the $L^{2}$ norm of the displacement error at the end of the computation. As can be seen, the use of the proposed projection technique has little effect on the convergence rate. We would like to point out that this is a purely elastic compressible problem, and hence the use of the proposed $\overline{\mathbf{F}}$ technique is not necessary. However, due to the availability of the exact solution, it provides useful insight into how the developed framework affects the convergence rates. 

\subsection{Cook's membrane}

\begin{figure}[!htbp]\centering
\includegraphics[width=0.4\textwidth]{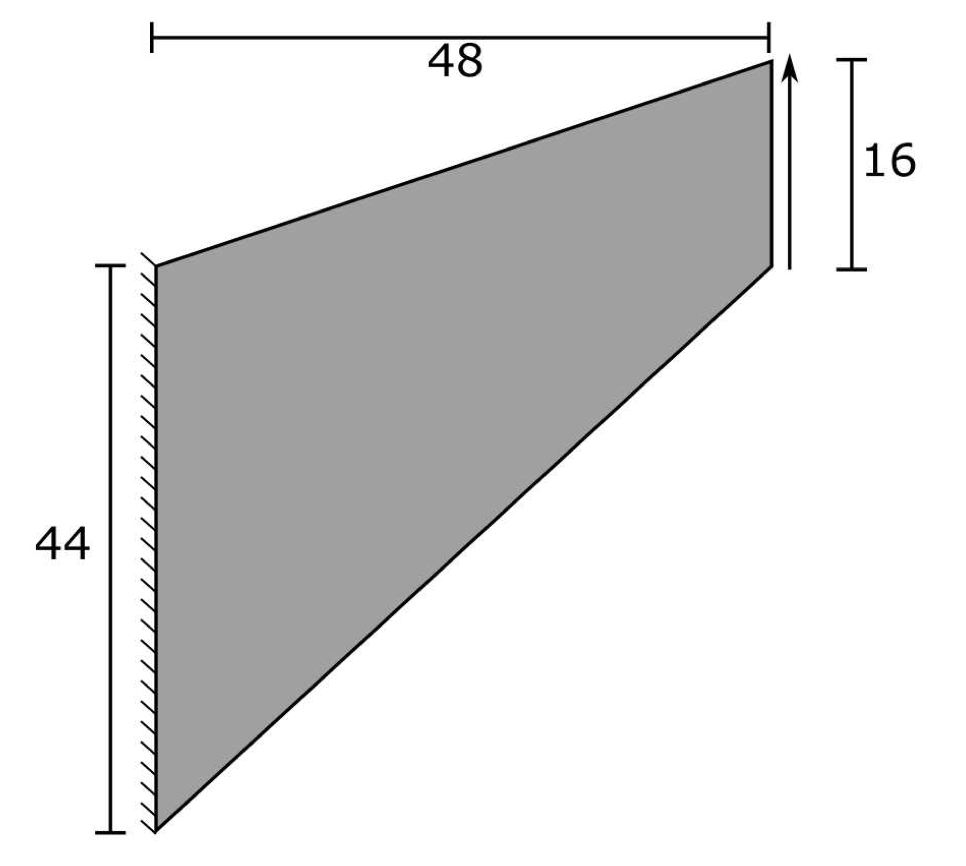}
\caption{Cook's membrane. The left edge is fixed and vertical traction is applied on the right edge.}
\label{fig:cookfigure}
\end{figure}

We now apply our proposed methodology to investigate the Cook's membrane problem. This problem is a traditional benchmark for studying incompressibility and has been examined by several researchers within the context of various numerical methods \cite{elguedj2008b,ostien201610,iaconeta2019stabilized,nakshatrala2008finite,moutsanidis2020treatment,auricchio2005analysis}. Here, we employ the exact setup that was used in \cite{moutsanidis2020treatment} for studying near-incompressibility within the context of immersed methods. An elastic model is used with Young's modulus $E=1000$~Pa, density $\rho = 1$~kg/$m^3$, and Poisson's ratio $\nu=0.499$. The setup of the problem can be seen in Figure \ref{fig:cookfigure} (dimensions in m). The left surface is fixed and a constant vertical traction of $0.25$~N/m is applied on the right surface. The background domain has dimensions $49$~m $\times$ $61$~m $\times$ $1$~m and we consider three discretizations, labeled M$N$, at refinement levels $N=1,2,3$. M1 has $25 \times 31 \times 1$ background elements and $6,710$ material points.  M2 has $50 \times 62 \times 1$ background elements and $26,900$ material points. M3 has $100 \times 124 \times 1$ background elements and $107,534$ material points. The above-mentioned discretizations correspond to 3 material points per background cell per $xy$-plane direction, whereas 2 material points are employed in the out of plane $z$-direction. The problem is solved with a three dimensional MPM code and plane strain conditions are enforced by restraining any out of plane action. Seven different combinations of background shape function order and $\overline{\mathbf{F}}$ projection are considered: 
\begin{enumerate}
\item $C^0$-continuous linear shape functions 
\item $C^1$-continuous quadratic B-spline shape functions
\item $C^2$-continuous cubic B-spline shape functions
\item $C^0$-continuous linear shape functions with $\overline{\mathbf{F}}$ projection onto a space of constants
\item $C^1$-continuous quadratic B-spline shape functions with $\overline{\mathbf{F}}$ projection onto a space of constants
\item $C^1$-continuous quadratic B-spline shape functions with $\overline{\mathbf{F}}$ projection onto a space of linears
\item $C^2$-continuous cubic B-spline shape functions with $\overline{\mathbf{F}}$ projection onto a space of quadratics
\end{enumerate}
The time step used with discretization M$N$ is $\Delta t = 2 \times 10^{-4}\times 2^{1-N}$~s and the final time is $3$~s.

\begin{figure}[!htbp]\centering
\includegraphics[width=0.5\textwidth]{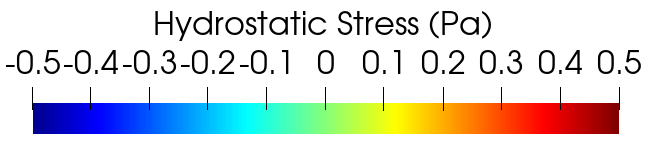}
\includegraphics[width=0.6\textwidth]{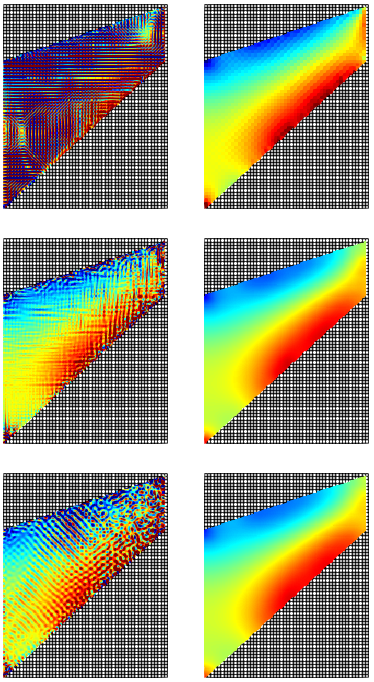}
\caption{Cook's membrane. Hydrostatic stress at the end of the computation. Discretization M2. Left column from top to bottom: $C^0$-continuous linear shape functions; $C^1$-continuous quadratic B-spline shape functions; $C^2$-continuous cubic B-spline shape functions. Right column from top to bottom: $C^0$-continuous linear shape functions with $\overline{\mathbf{F}}$ projection onto constants; $C^1$-continuous quadratic B-spline shape functions with $\overline{\mathbf{F}}$ projection onto linears; $C^2$-continuous cubic B-spline shape functions with $\overline{\mathbf{F}}$ projection onto quadratics.}
\label{fig:cookplot1}
\end{figure}

\begin{figure}[!htbp]\centering
\includegraphics[width=0.5\textwidth]{Figures/cookcolormap.png}
\includegraphics[width=0.6\textwidth]{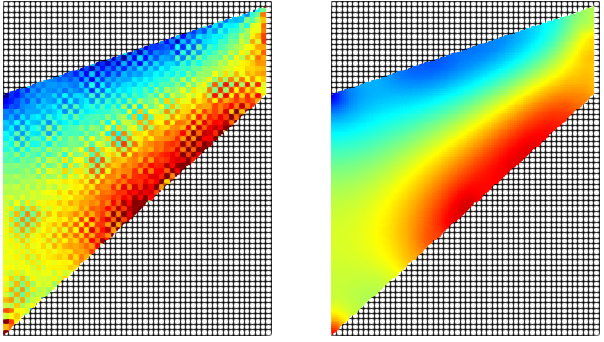}
\caption{Cook's membrane. Hydrostatic stress at the end of the computation. Discretization M2. Left: $C^1$-continuous quadratic B-spline shape functions with $\overline{\mathbf{F}}$ projection onto constants. Right: $C^1$-continuous quadratic B-spline shape functions with $\overline{\mathbf{F}}$ projection onto linears.}
\label{fig:cookplot2}
\end{figure}

\begin{figure}[!htbp]\centering
\includegraphics[width=0.5\textwidth]{Figures/cookcolormap.png}
\includegraphics[width=0.7\textwidth]{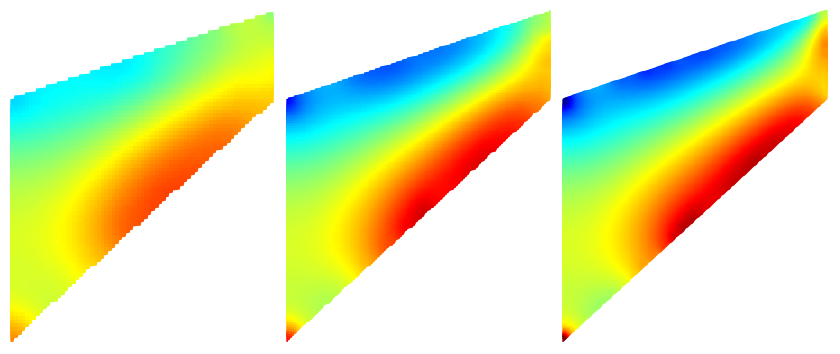}
\caption{Cook's membrane. Hydrostatic stress at the end of the computation. $C^1$-continuous quadratic B-spline shape functions with $\overline{\mathbf{F}}$ projection onto linears. From left to right: M1; M2; M3.}
\label{fig:cookplot3}
\end{figure}

\begin{figure}[!htbp]\centering
\includegraphics[width=0.9\textwidth]{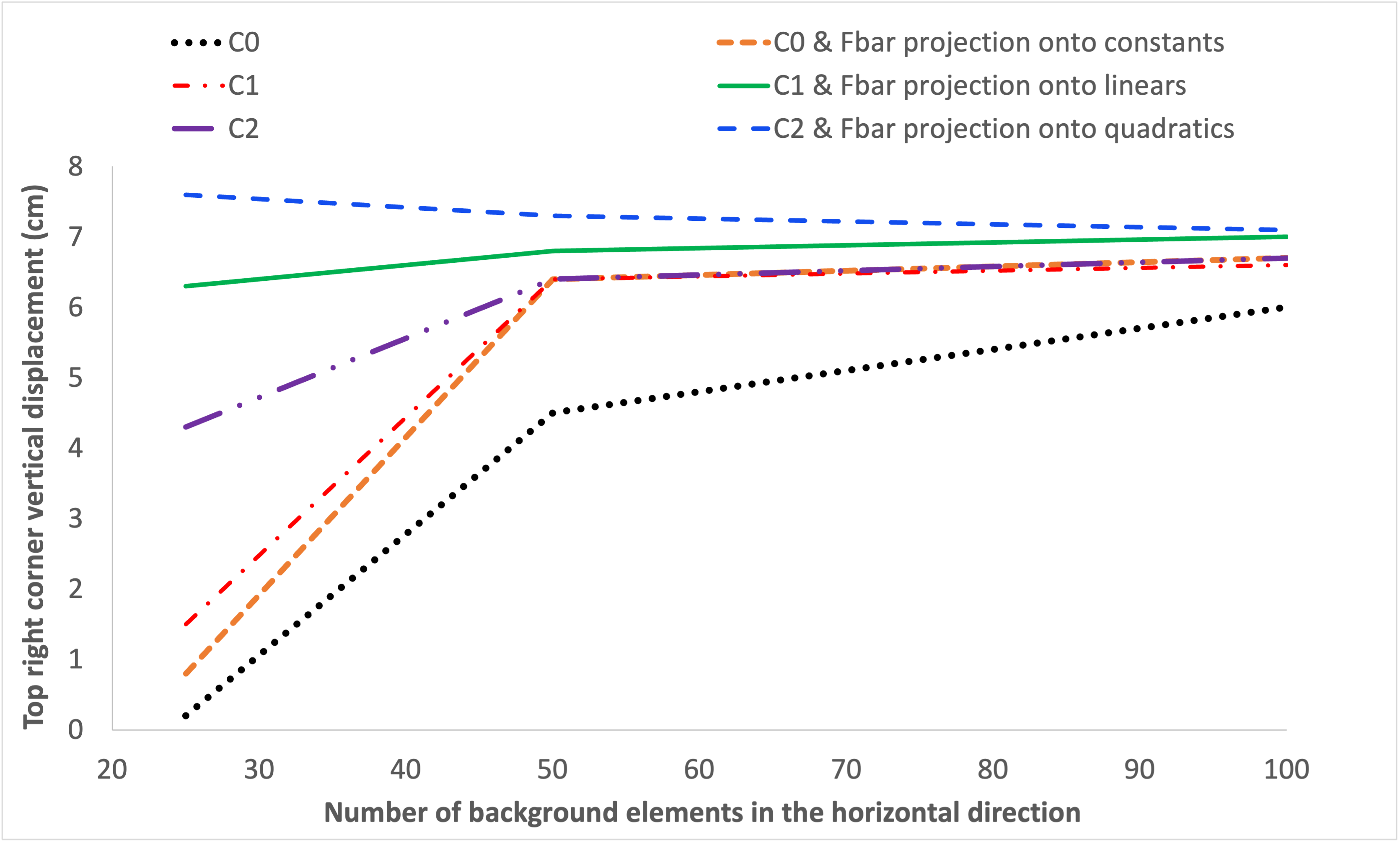}
\caption{Cook's membrane. Convergence of top right corner vertical displacement under mesh refinement.}
\label{fig:cookplot4}
\end{figure}

Snapshots of the hydrostatic stress field for discretization M2 at the end of the computation are provided in Figure \ref{fig:cookplot1}. It can be seen that our proposed projection method results in a smooth stress field and reduced stress oscillations. It can also be observed that the use of higher order spline shape functions alone is not sufficient to fix the stress oscillation issue and needs to be combined with our proposed framework. Furthermore, the hydrostatic stress contours are in excellent agreement with \cite{moutsanidis2020treatment}. In Figure \ref{fig:cookplot2} we demonstrate the effect of the projection space order on the hydrostatic stress for the case of $C^{1}$-continuous quadratic B-spline shape functions. As can be seen, the $\overline{\mathbf{F}}$ projection onto linears produces a much smoother stress field than the projection onto constants. This is consistent with the spirit of the original $\overline{\mathbf{F}}$ method in which the projection space is one order lower than the approximation space. In Figure \ref{fig:cookplot3} we demonstrate that the proposed projection leads to convergent pressure contours under refinement. In Figure \ref{fig:cookplot4} we plot the convergence of the top right corner vertical displacement. It is observed that the traditional MPM ($C^0$-continuous linear background shape functions) exhibits volumetric locking. The results are improved as the approximation order is elevated and by using our proposed $\overline{\mathbf{F}}$ projection. Clearly, the proposed methodology overcomes volumetric locking, and the results are in good agreement with the results obtained in \cite{moutsanidis2020treatment}. 

\subsection{Elasto-plastic collapse}

This example is similar to the elasto-plastic collapse presented in \cite{coombs2018overcoming}, in which the collapse of a $16$~m $\times$ $8$~m plane strain block under the application of a body force is studied. Due to symmetry, a block of $8$~m $\times$ $8$~m $\times$ $1$~m is analyzed instead, and roller boundary conditions are applied along the line of symmetry. The background domain has dimensions $15$~m $\times$ $10$~m $\times$ $1$~m and is discretized with $30 \times 20 \times 1$ elements. The block is discretized with 5 particles per background cell per $xy$-plane direction, whereas 2 material points are used in the out of plane $z$-direction. The problem is solved with a three dimensional MPM code and plane strain conditions are enforced by restraining any out of plane action. J$_2$ plasticity with no hardening is considered, with Young's modulus $E = 100$~kPa, density $\rho = 1$~kg/$m^3$, Poisson ratio $\nu = 0.3$, and yield stress $\sigma_{y} = 15$~kPa. The block is subjected to a body force of $-3$~kN/$m^3$. The analysis is run for $0.3$~s with a time step of $5 \times 10^{-4}$~s. For the background discretization we employ $C^0$-continuous linear, $C^1$-continuous quadratic, and $C^2$-continuous cubic B-spline shape functions. For all three of them we perform analyses with and without the proposed $\overline{\mathbf{F}}$ projection. 

\begin{figure}[!htbp]\centering
\includegraphics[width=0.6\textwidth]{Figures/Colormap1.png}
\includegraphics[width=1.0\textwidth]{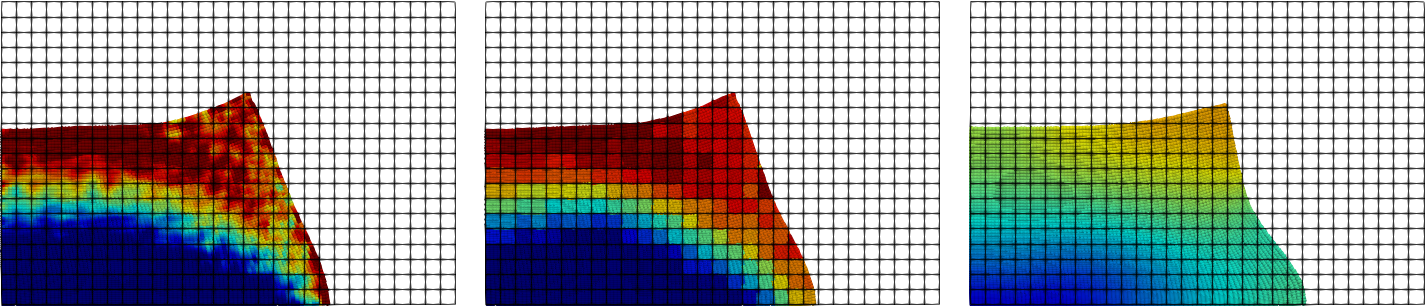}
\caption{Elasto-plastic collapse. Hydrostatic stress at the end of the computation. From left to right: $C^1$-continuous quadratic B-spline shape functions; $C^1$-continuous quadratic B-spline shape functions with $\overline{\mathbf{F}}$ projection onto constants; $C^1$-continuous quadratic B-spline shape functions with $\overline{\mathbf{F}}$ projection onto linears.}
\label{fig:Pressure plot for quadratics}
\end{figure}

The hydrostatic stress contours at the end of the computation are presented in Figure \ref{fig:Pressure plot for linears, quad and cubics}. As can be seen, the use of $\overline{\mathbf{F}}$ projection results in much smoother hydrostatic stress field with significantly reduced oscillations. This is particularly pronounced in the case of higher order MPM ($C^1$-continuous quadratic and $C^2$-continuous cubic B-spline shape functions) where it can be seen that the use of higher order shape functions alone is not enough to remove the pressure oscillations. We also study the effect of using higher order MPM and employing an $\overline{\mathbf{F}}$ projection onto a space of constant functions (Figure \ref{fig:Pressure plot for quadratics}). It can be seen that although using an $\overline{\mathbf{F}}$ projection onto constants results in smoother hydrostatic stress compared to not using the projection at all, the results can be further improved by using an $\overline{\mathbf{F}}$ projection onto a space of linear functions. Finally, it can be observed that employing $\overline{\mathbf{F}}$ results in slightly higher displacements, and thus alleviates volumetric locking.

\begin{figure}[!htbp]\centering
\includegraphics[width=0.6\textwidth]{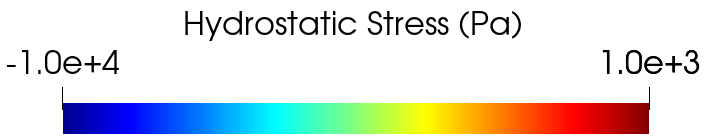}
\includegraphics[width=0.8\textwidth]{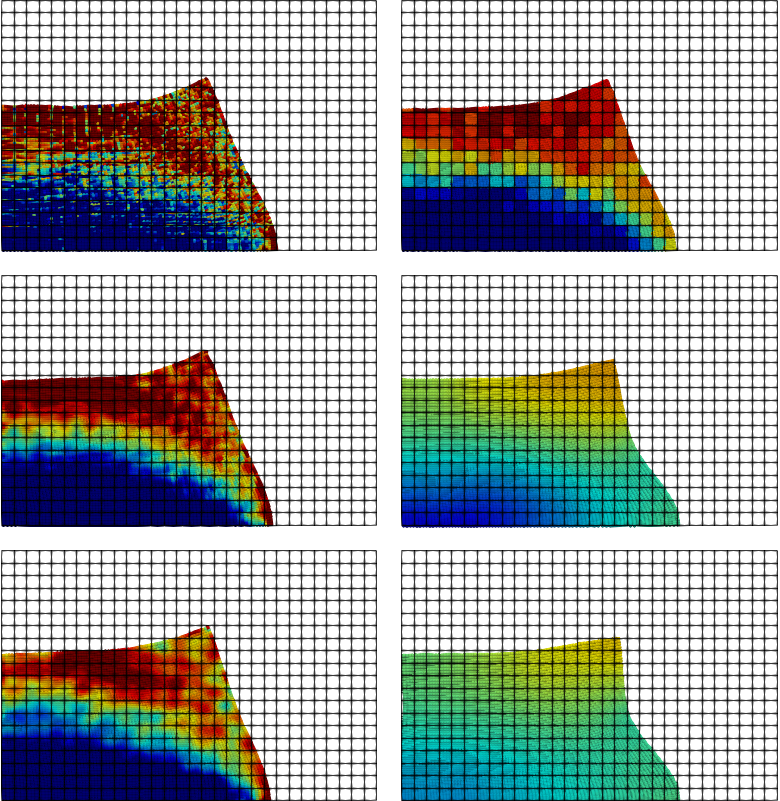}
\caption{Elasto-plastic collapse. Hydrostatic stress at the end of the computation. Left column from top to bottom: $C^0$-continuous linear shape functions; $C^1$-continuous quadratic B-spline shape functions; $C^2$-continuous cubic B-spline shape functions. Right column from top to bottom: $C^0$-continuous linear shape functions with $\overline{\mathbf{F}}$ projection onto constants; $C^1$-continuous quadratic B-spline shape functions with $\overline{\mathbf{F}}$ projection onto linears; $C^2$-continuous cubic B-spline shape functions with $\overline{\mathbf{F}}$ projection onto quadratics.}
\label{fig:Pressure plot for linears, quad and cubics}
\end{figure}

\subsection{Taylor bar impact}

\begin{figure}[!htbp]\centering
\includegraphics[width=0.6\textwidth]{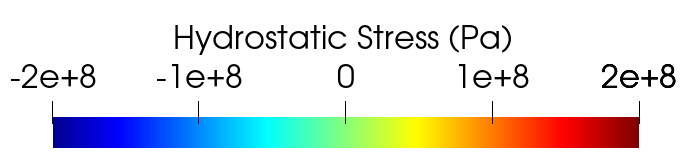}
  \begin{tabular}{cccc}
    \includegraphics[width=0.2\textwidth]{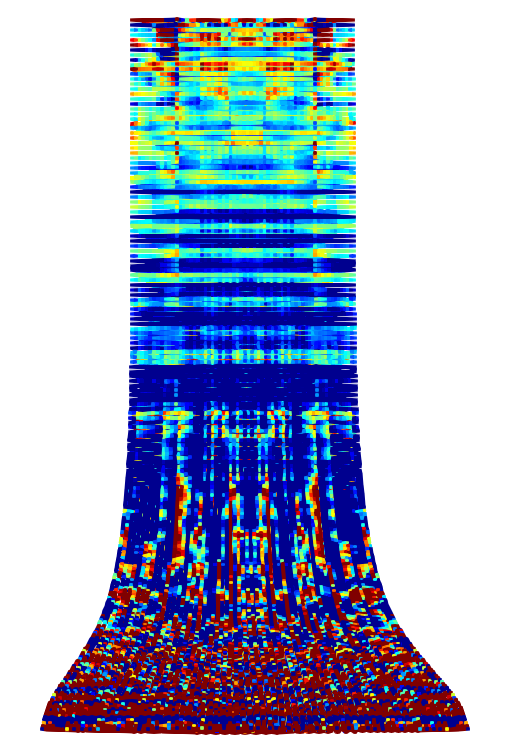}&\includegraphics[width=0.2\textwidth]{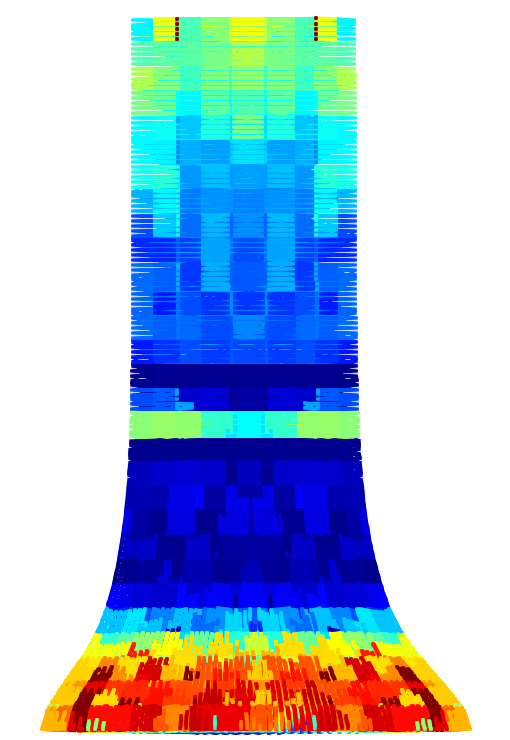}&\includegraphics[width=0.2\textwidth]{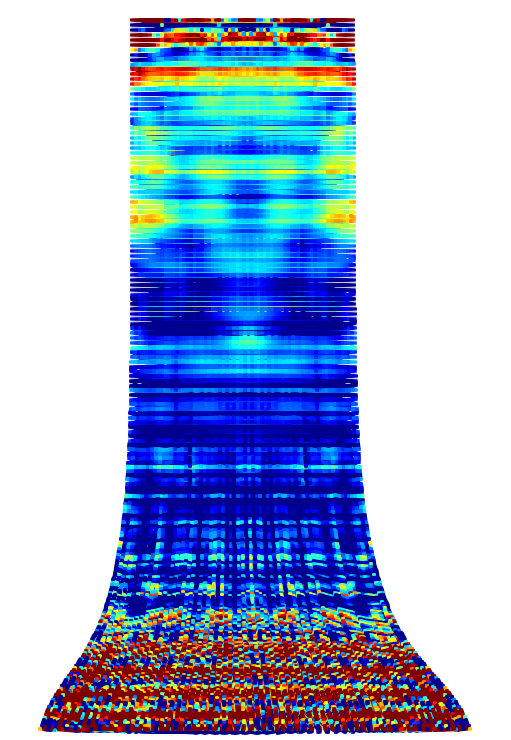}&\includegraphics[width=0.2\textwidth]{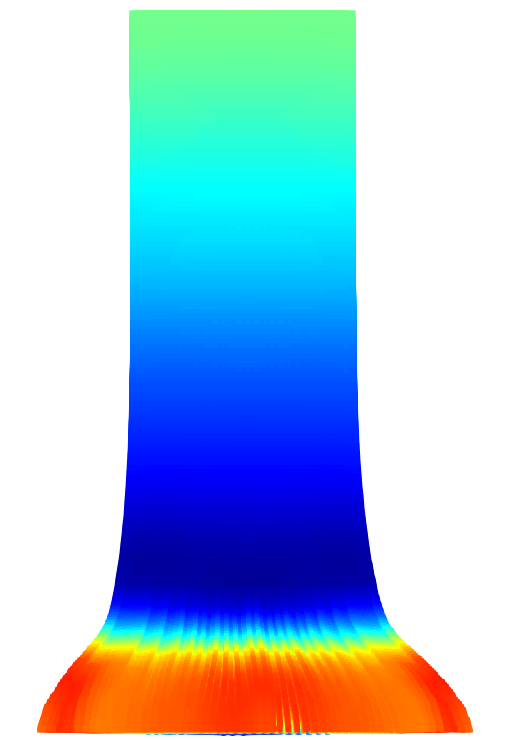}\\
\end{tabular}
\caption{Taylor bar impact. Hydrostatic stress. View at intermediate configuration. From left to right: $C^{0}$-continuous linear shape functions; $C^{0}$-continuous linear shape functions with $\overline{\mathbf{F}}$ projection onto constants; $C^{1}$-continuous quadratic B-spline shape functions; $C^{1}$-continuous quadratic B-spline shape functions with $\overline{\mathbf{F}}$ projection onto linears.}
\label{fig:taylor_intermediate}
\end{figure}

\begin{figure}[!htbp]\centering
\includegraphics[width=0.6\textwidth]{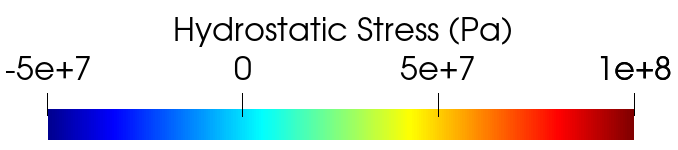}
  \begin{tabular}{cccc}
    \includegraphics[width=0.2\textwidth]{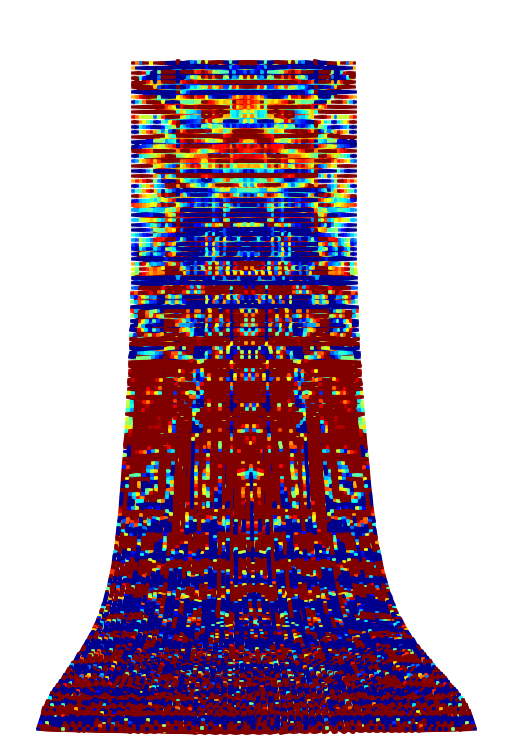}&\includegraphics[width=0.2\textwidth]{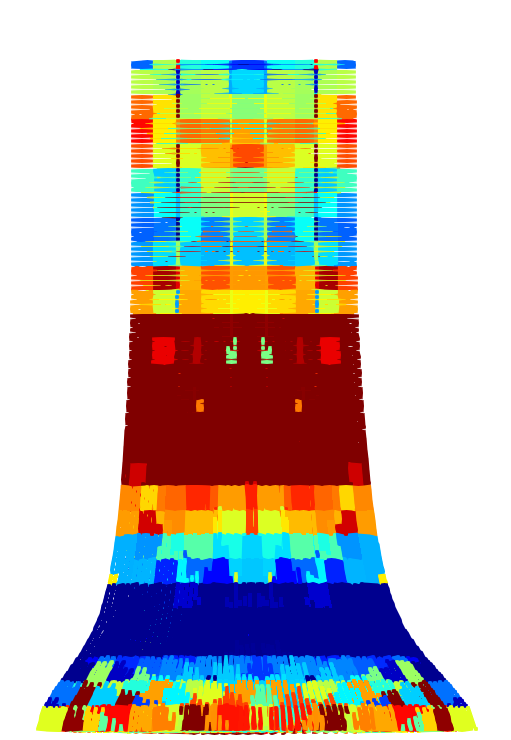}&\includegraphics[width=0.2\textwidth]{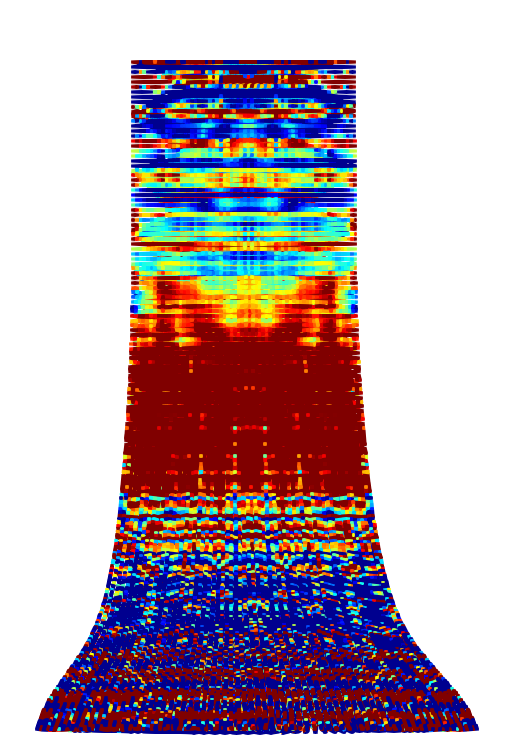}&\includegraphics[width=0.2\textwidth]{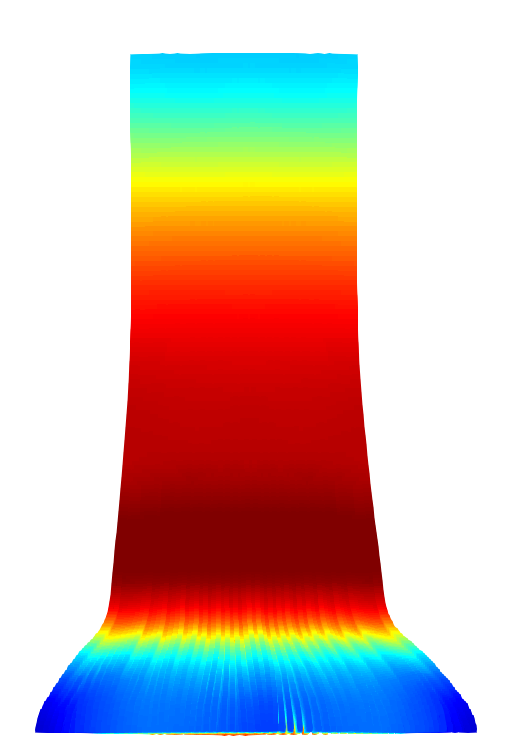}\\
\end{tabular}
\caption{Taylor bar impact.  Hydrostatic stress. View at final configuration. From left to right: $C^{0}$-continuous linear shape functions; $C^{0}$-continuous linear shape functions with $\overline{\mathbf{F}}$ projection onto constants; $C^{1}$-continuous quadratic B-spline shape functions; $C^{1}$-continuous quadratic B-spline shape functions with $\overline{\mathbf{F}}$ projection onto linears.}
\label{fig:taylor_final}
\end{figure}

We finally apply our formulation to a 3D Taylor bar impact problem, and we employ the specific setup used in \cite{hillman2016accelerated,moutsanidis2020iga}. We consider a cylindrical aluminum bar with an initial height of $2.346$~cm and radius of $0.391$~cm, that impacts a rigid wall with an initial velocity of $373$~m/s. $J_{2}$ plasticity with isotropic hardening is considered, and the material properties are Young's modulus $E = 78.2$~GPa, Poisson's ratio $\nu = 0.30$, and density $\rho = 2700$~kg/m$^3$. The yield function is taken as
\begin{equation}
f (\pmb{\sigma} \, ^{dev},\overline{e}_{p}) = || \pmb{\sigma}^{dev} || - \sqrt{\frac{2}{3}} K (\overline{e}_{p}),
\end{equation}
where
\begin{equation}
K (\overline{e}_{p}) = \sigma_{Y} (1+125 \, \overline{e}_{p})^{0.1},
\end{equation}
$\overline{e}_{p}$ is the equivalent plastic strain, and $\sigma_{Y} = 0.29$~GPa. Due to the doubly symmetric nature of the problem, quarter symmetry is considered. The background domain has dimensions $1.2 \times 1.2 \times 2.4$~cm$^3$. The background domain is discretized with $20 \times 20 \times 40$ elements and the bar is discretized with $162,876$ particles. The time step used is $\Delta t = 10^{-8}$~s and the final time is $4 \times 10^{-5}$~s. The dimensions of the bar at the final configuration are shown in Table \ref{table:1}. For comparison purposes we provide the experimental results and the FEM results \cite{wilkins1973impact}, as well as the results from computations performed with the reproducing kernel particle method (RKPM) \cite{hillman2016accelerated}, and the particle-in-cell method \cite{sulsky1995application}. It can be observed that combining higher order shape functions with our proposed $\overline{\mathbf{F}}$ projection eliminates volumetric locking and the results are in excellent agreement with the published literature. The hydrostatic stress contours at an intermediate configuration are provided in Figure \ref{fig:taylor_intermediate}, and at the final configuration in Figure \ref{fig:taylor_final}. As can be seen, our proposed projection method produces much smoother stress fields, while maintaining all the major qualitative and quantitative features of the original Taylor bar analysis.

\begin{table}[!htbp]
\caption{Dimensions of deformed 3D Taylor bar}
\centering
\begin{tabular}{c|c|c}
\hline
& Radius (cm) & Height (cm) \\
\hline
$C^{0}$-continuous & 0.768 & 1.632\\
\hline
$C^{1}$-continuous & 0.776 & 1.634\\
\hline
$C^{0}$-continuous with $\overline{\mathbf{F}}$ projection onto constants & 0.77 & 1.634\\
\hline
$C^{1}$-continuous with $\overline{\mathbf{F}}$ projection onto linears & 0.776 & 1.649\\
\hline
RKPM \cite{hillman2016accelerated}  & 0.775 & 1.651\\
\hline
PIC \cite{sulsky1995application}  & 0.78 & 1.65\\
\hline
FEM \cite{wilkins1973impact} & 0.742 & 1.652 \\
\hline
Experimental \cite{wilkins1973impact} & - & 1.651 \\
\hline
\end{tabular}
\label{table:1}
\end{table}

\section{Conclusions}\label{sec:conclusions}

We presented a new $\overline{\mathbf{F}}$ projection technique for near-incompressible large deformation elasticity and plasticity within the context of higher order material point methods. The proposed approach is based on the projection of the dilatational (volumetric) part of the velocity gradient onto a lower dimensional approximation space. Specifically, an improved velocity gradient was derived based on the projection of its dilatational part onto a space whose order is reduced by one compared to the approximation space of trial and test functions. A collection of challenging examples was presented and the proposed methodology exhibited reduced stress oscillations, and was free of volumetric locking and low energy modes.

\section*{Acknowledgments}
The computations were carried out in PetIGA, a software framework that implements NURBS-based IGA~\cite{dalcin2016petiga}. The authors would like to thank Stony Brook Research Computing and Cyberinfrastructure, and the Institute for Advanced Computational Science at Stony Brook University for access to the high-performance SeaWulf computing system, which was made possible by a \$1.4M National Science Foundation grant (\#1531492).

\bibliographystyle{unsrt}
\bibliography{main}

\begin{thebibliography}{10}

\bibitem{sulsky1994particle}
Deborah Sulsky, Zhen Chen, and Howard~L Schreyer.
\newblock A particle method for history-dependent materials.
\newblock {\em Computer Methods in Applied Mechanics and Engineering},
  118(1-2):179--196, 1994.

\bibitem{harlow1962particle}
Francis~H Harlow.
\newblock The particle-in-cell method for numerical solution of problems in
  fluid dynamics.
\newblock Technical report, Los Alamos Scientific Lab., N. Mex., 1962.

\bibitem{evans1957particle}
Martha~W Evans, Francis~H Harlow, and Eleazer Bromberg.
\newblock The particle-in-cell method for hydrodynamic calculations.
\newblock Technical report, LOS ALAMOS NATIONAL LAB NM, 1957.

\bibitem{hughes2012finite}
Thomas~JR Hughes.
\newblock {\em The finite element method: linear static and dynamic finite
  element analysis}.
\newblock Courier Corporation, 2012.

\bibitem{sulsky1993particle}
D~Sulsky and HL~Schreyer.
\newblock A particle method with large rotations applied to the penetration of
  history-dependent materials.
\newblock {\em ASME APPLIED MECHANICS DIVISION-PUBLICATIONS-AMD}, 171:95--95,
  1993.

\bibitem{sulsky1993particle1}
D~Sulsky and HL~Schreyer.
\newblock The particle-in-cell method as a natural impact algorithm.
\newblock {\em ASME APPLIED MECHANICS DIVISION-PUBLICATIONS-AMD}, 180:219--219,
  1993.

\bibitem{york2000fluid}
Allen~R York, Deborah Sulsky, and Howard~L Schreyer.
\newblock Fluid--membrane interaction based on the material point method.
\newblock {\em International Journal for Numerical Methods in Engineering},
  48(6):901--924, 2000.

\bibitem{stomakhin2013material}
Alexey Stomakhin, Craig Schroeder, Lawrence Chai, Joseph Teran, and Andrew
  Selle.
\newblock A material point method for snow simulation.
\newblock {\em ACM Transactions on Graphics (TOG)}, 32(4):102, 2013.

\bibitem{andersen2010modelling}
S{\o}ren Andersen and Lars Andersen.
\newblock Modelling of landslides with the material-point method.
\newblock {\em Computational Geosciences}, 14(1):137--147, 2010.

\bibitem{bardenhagen2004generalized}
SG~Bardenhagen and EM~Kober.
\newblock The generalized interpolation material point method.
\newblock {\em Computer Modeling in Engineering and Sciences}, 5(6):477--496,
  2004.

\bibitem{chen2002evaluation}
Zhen Chen and Rebecca Brannon.
\newblock An evaluation of the material point method.
\newblock {\em SAND Report, SAND2002-0482,(February 2002)}, 2002.

\bibitem{zhang2008material}
Duan~Z Zhang, Qisu Zou, W~Brian VanderHeyden, and Xia Ma.
\newblock Material point method applied to multiphase flows.
\newblock {\em Journal of Computational Physics}, 227(6):3159--3173, 2008.

\bibitem{zhang2011material}
Duan~Z Zhang, Xia Ma, and Paul~T Giguere.
\newblock Material point method enhanced by modified gradient of shape
  function.
\newblock {\em Journal of Computational Physics}, 230(16):6379--6398, 2011.

\bibitem{sadeghirad2011convected}
A~Sadeghirad, Rebecca~M Brannon, and J~Burghardt.
\newblock A convected particle domain interpolation technique to extend
  applicability of the material point method for problems involving massive
  deformations.
\newblock {\em International Journal for Numerical Methods in Engineering},
  86(12):1435--1456, 2011.

\bibitem{yerro2015material}
A~Yerro, EE~Alonso, and NM~Pinyol.
\newblock The material point method for unsaturated soils.
\newblock {\em G{\'e}otechnique}, 65(3):201--217, 2015.

\bibitem{soga2016trends}
Kenichi Soga, E~Alonso, Alba Yerro, K~Kumar, and Samila Bandara.
\newblock Trends in large-deformation analysis of landslide mass movements with
  particular emphasis on the material point method.
\newblock {\em G{\'e}otechnique}, 66(3):248--273, 2016.

\bibitem{homel2017field}
Michael~A Homel and Eric~B Herbold.
\newblock Field-gradient partitioning for fracture and frictional contact in
  the material point method.
\newblock {\em International Journal for Numerical Methods in Engineering},
  109(7):1013--1044, 2017.

\bibitem{kumar2017modelling}
Krishna Kumar, Kenichi Soga, Jean-Yves Delenne, and Farhang Radjai.
\newblock Modelling transient dynamics of granular slopes: {MPM} and {DEM}.
\newblock {\em Procedia Engineering}, 175:94--101, 2017.

\bibitem{tielen2017high}
Roel Tielen, Elizaveta Wobbes, Matthias M{\"o}ller, and Lars Beuth.
\newblock A high order material point method.
\newblock {\em Procedia Engineering}, 175:265--272, 2017.

\bibitem{gan2018enhancement}
Yong Gan, Zheng Sun, Zhen Chen, Xiong Zhang, and Yu~Liu.
\newblock Enhancement of the material point method using b-spline basis
  functions.
\newblock {\em International Journal for Numerical Methods in Engineering},
  113(3):411--431, 2018.

\bibitem{moutsanidis2019modeling}
Georgios Moutsanidis, David Kamensky, Duan~Z Zhang, Yuri Bazilevs, and
  Christopher~C Long.
\newblock Modeling strong discontinuities in the material point method using a
  single velocity field.
\newblock {\em Computer Methods in Applied Mechanics and Engineering},
  345:584--601, 2019.

\bibitem{kumar2019scalable}
Krishna Kumar, Jeffrey Salmond, Shyamini Kularathna, Christopher Wilkes, Ezra
  Tjung, Giovanna Biscontin, and Kenichi Soga.
\newblock Scalable and modular material point method for large-scale
  simulations.
\newblock {\em arXiv preprint arXiv:1909.13380}, 2019.

\bibitem{moutsanidis2020iga}
Georgios Moutsanidis, Christopher~C Long, and Yuri Bazilevs.
\newblock {IGA}-{MPM}: The isogeometric material point method.
\newblock {\em Computer Methods in Applied Mechanics and Engineering},
  372:113346, 2020.

\bibitem{de2020material}
Alban de~Vaucorbeil, Vinh~Phu Nguyen, Sina Sinaie, and Jian~Ying Wu.
\newblock Material point method after 25 years: Theory, implementation, and
  applications.
\newblock {\em Advances in Applied Mechanics}, 53:185--398, 2020.

\bibitem{chen2018vp}
Zhen-Peng Chen, Xiong Zhang, Kam~Yim Sze, Lei Kan, and Xin-Ming Qiu.
\newblock vp material point method for weakly compressible problems.
\newblock {\em Computers \& Fluids}, 176:170--181, 2018.

\bibitem{hughes1977equivalence}
Thomas~JR Hughes.
\newblock Equivalence of finite elements for nearly incompressible elasticity.
\newblock 1977.

\bibitem{malkus1978mixed}
David~S Malkus and Thomas~JR Hughes.
\newblock Mixed finite element methods—reduced and selective integration
  techniques: a unification of concepts.
\newblock {\em Computer Methods in Applied Mechanics and Engineering},
  15(1):63--81, 1978.

\bibitem{hughes1980generalization}
Thomas~JR Hughes.
\newblock Generalization of selective integration procedures to anisotropic and
  nonlinear media.
\newblock {\em International Journal for Numerical Methods in Engineering},
  15(9):1413--1418, 1980.

\bibitem{belytschko1984hourglass}
Ted Belytschko, Jame Shau-Jen Ong, Wing~Kam Liu, and James~M Kennedy.
\newblock Hourglass control in linear and nonlinear problems.
\newblock {\em Computer Methods in Applied Mechanics and Engineering},
  43(3):251--276, 1984.

\bibitem{simo1990class}
Juan~C Simo and MS~Rifai.
\newblock A class of mixed assumed strain methods and the method of
  incompatible modes.
\newblock {\em International Journal for Numerical Methods in Engineering},
  29(8):1595--1638, 1990.

\bibitem{neto996design}
EA~de~Souza~Neto, D~Peri{\'c}, M~Dutko, and DRJ Owen.
\newblock Design of simple low order finite elements for large strain analysis
  of nearly incompressible solids.
\newblock {\em International Journal of Solids and Structures},
  33(20-22):3277--3296, 1996.

\bibitem{klaas1999stabilized}
Ottmar Klaas, Antoinette Maniatty, and Mark~S Shephard.
\newblock A stabilized mixed finite element method for finite elasticity.:
  Formulation for linear displacement and pressure interpolation.
\newblock {\em Computer Methods in Applied Mechanics and Engineering},
  180(1-2):65--79, 1999.

\bibitem{kasper2000mixed}
Eric~P Kasper and Robert~L Taylor.
\newblock A mixed-enhanced strain method: Part {I}: Geometrically linear
  problems.
\newblock {\em Computers \& Structures}, 75(3):237--250, 2000.

\bibitem{neto2005f}
EA~De~Souza Neto, FM~Andrade Pires, and DRJ Owen.
\newblock F-bar-based linear triangles and tetrahedra for finite strain
  analysis of nearly incompressible solids. {P}art {I}: formulation and
  benchmarking.
\newblock {\em International Journal for Numerical Methods in Engineering},
  62(3):353--383, 2005.

\bibitem{elguedj2008b}
Thomas Elguedj, Yuri Bazilevs, Victor~M Calo, and Thomas~JR Hughes.
\newblock B and {F} projection methods for nearly incompressible linear and
  non-linear elasticity and plasticity using higher-order nurbs elements.
\newblock {\em Computer Methods in Applied Mechanics and Engineering},
  197(33-40):2732--2762, 2008.

\bibitem{brezzi2012mixed}
Franco Brezzi and Michel Fortin.
\newblock {\em Mixed and hybrid finite element methods}, volume~15.
\newblock Springer Science \& Business Media, 2012.

\bibitem{zhang2017incompressible}
Fan Zhang, Xiong Zhang, Kam~Yim Sze, Yanping Lian, and Yan Liu.
\newblock Incompressible material point method for free surface flow.
\newblock {\em Journal of Computational Physics}, 330:92--110, 2017.

\bibitem{kularathna2017implicit}
Shyamini Kularathna and Kenichi Soga.
\newblock Implicit formulation of material point method for analysis of
  incompressible materials.
\newblock {\em Computer Methods in Applied Mechanics and Engineering},
  313:673--686, 2017.

\bibitem{zhang2018augmented}
Fan Zhang, Xiong Zhang, and Yan Liu.
\newblock An augmented incompressible material point method for modeling liquid
  sloshing problems.
\newblock {\em International Journal of Mechanics and Materials in Design},
  14(1):141--155, 2018.

\bibitem{love2006energy}
E~Love and Deborah~L Sulsky.
\newblock An energy-consistent material-point method for dynamic finite
  deformation plasticity.
\newblock {\em International Journal for Numerical Methods in Engineering},
  65(10):1608--1638, 2006.

\bibitem{mast2012mitigating}
CM~Mast, P~Mackenzie-Helnwein, Pedro Arduino, Gregory~R Miller, and W~Shin.
\newblock Mitigating kinematic locking in the material point method.
\newblock {\em Journal of Computational Physics}, 231(16):5351--5373, 2012.

\bibitem{iaconeta2019stabilized}
Ilaria Iaconeta, Antonia Larese, Riccardo Rossi, and Eugenio O{\~n}ate.
\newblock A stabilized mixed implicit material point method for non-linear
  incompressible solid mechanics.
\newblock {\em Computational Mechanics}, 63(6):1243--1260, 2019.

\bibitem{coombs2018overcoming}
William~M Coombs, Tim~J Charlton, Michael Cortis, and Charles~E Augarde.
\newblock Overcoming volumetric locking in material point methods.
\newblock {\em Computer Methods in Applied Mechanics and Engineering},
  333:1--21, 2018.

\bibitem{hughes2005isogeometric}
Thomas~JR Hughes, John~A Cottrell, and Yuri Bazilevs.
\newblock Isogeometric analysis: Cad, finite elements, nurbs, exact geometry
  and mesh refinement.
\newblock {\em Computer Methods in Applied Mechanics and Engineering},
  194(39-41):4135--4195, 2005.

\bibitem{steffen2008analysis}
Michael Steffen, Robert~M Kirby, and Martin Berzins.
\newblock Analysis and reduction of quadrature errors in the material point
  method (mpm).
\newblock {\em International Journal for Numerical Methods in Engineering},
  76(6):922--948, 2008.

\bibitem{de2021extension}
Pascal de~Koster, Roel Tielen, Elizaveta Wobbes, and Matthias M{\"o}ller.
\newblock Extension of b-spline {M}aterial {P}oint {M}ethod for unstructured
  triangular grids using {P}owell--{S}abin splines.
\newblock {\em Computational Particle Mechanics}, 8(2):273--288, 2021.

\bibitem{bazilevs2017new}
Yuri Bazilevs, Georgios Moutsanidis, Jesus Bueno, Kazem Kamran, David Kamensky,
  Michael~Charles Hillman, Hector Gomez, and JS~Chen.
\newblock A new formulation for air-blast fluid--structure interaction using an
  immersed approach: part {II}—coupling of {IGA} and meshfree
  discretizations.
\newblock {\em Computational Mechanics}, 60(1):101--116, 2017.

\bibitem{moutsanidis2018hyperbolic}
Georgios Moutsanidis, David Kamensky, JS~Chen, and Yuri Bazilevs.
\newblock Hyperbolic phase field modeling of brittle fracture: Part
  {II}—immersed {IGA}--{RKPM} coupling for air-blast--structure interaction.
\newblock {\em Journal of the Mechanics and Physics of Solids}, 121:114--132,
  2018.

\bibitem{bazilevs2006isogeometric1}
Yuri Bazilevs, L~Beirao~da Veiga, J~Austin Cottrell, Thomas~JR Hughes, and
  Giancarlo Sangalli.
\newblock Isogeometric analysis: approximation, stability and error estimates
  for h-refined meshes.
\newblock {\em Mathematical Models and Methods in Applied Sciences},
  16(07):1031--1090, 2006.

\bibitem{cox1972numerical}
Maurice~G Cox.
\newblock The numerical evaluation of b-splines.
\newblock {\em IMA Journal of Applied Mathematics}, 10(2):134--149, 1972.

\bibitem{simo2006computational}
Juan~C Simo and Thomas~JR Hughes.
\newblock {\em Computational Inelasticity}, volume~7.
\newblock Springer Science \& Business Media, 2006.

\bibitem{wobbes2021comparison}
Elizaveta Wobbes, Roel Tielen, Matthias M{\"o}ller, and Cornelis Vuik.
\newblock Comparison and unification of material-point and optimal
  transportation meshfree methods.
\newblock {\em Computational Particle Mechanics}, 8(1):113--133, 2021.

\bibitem{ostien201610}
Jakob~T Ostien, James~W Foulk, Alejandro Mota, and MG~Veilleux.
\newblock A 10-node composite tetrahedral finite element for solid mechanics.
\newblock {\em International Journal for Numerical Methods in Engineering},
  107(13):1145--1170, 2016.

\bibitem{nakshatrala2008finite}
KB~Nakshatrala, Arif Masud, and KD~Hjelmstad.
\newblock On finite element formulations for nearly incompressible linear
  elasticity.
\newblock {\em Computational Mechanics}, 41(4):547--561, 2008.

\bibitem{moutsanidis2020treatment}
Georgios Moutsanidis, Jacob~J Koester, Michael~R Tupek, Jiun-Shyan Chen, and
  Yuri Bazilevs.
\newblock Treatment of near-incompressibility in meshfree and immersed-particle
  methods.
\newblock {\em Computational Particle Mechanics}, 7(2):309--327, 2020.

\bibitem{auricchio2005analysis}
Ferdinando Auricchio, L~Beirao Da~Veiga, Carlo Lovadina, and Alessandro Reali.
\newblock An analysis of some mixed-enhanced finite element for plane linear
  elasticity.
\newblock {\em Computer Methods in Applied Mechanics and Engineering},
  194(27-29):2947--2968, 2005.

\bibitem{hillman2016accelerated}
Michael Hillman and Jiun-Shyan Chen.
\newblock An accelerated, convergent, and stable nodal integration in galerkin
  meshfree methods for linear and nonlinear mechanics.
\newblock {\em International Journal for Numerical Methods in Engineering},
  107(7):603--630, 2016.

\bibitem{wilkins1973impact}
Mark~L Wilkins and Michael~W Guinan.
\newblock Impact of cylinders on a rigid boundary.
\newblock {\em Journal of Applied Physics}, 44(3):1200--1206, 1973.

\bibitem{sulsky1995application}
Deborah Sulsky, Shi-Jian Zhou, and Howard~L Schreyer.
\newblock Application of a particle-in-cell method to solid mechanics.
\newblock {\em Computer Physics Communications}, 87(1-2):236--252, 1995.

\bibitem{dalcin2016petiga}
Lisandro Dalcin, Nathaniel Collier, Philippe Vignal, AMA C{\^o}rtes, and
  Victor~M Calo.
\newblock Petiga: A framework for high-performance isogeometric analysis.
\newblock {\em Computer Methods in Applied Mechanics and Engineering},
  308:151--181, 2016.

\end{thebibliography}

\end{document}